\documentclass[%
reprint,
 amsmath,amssymb,
 aps,
 prb,
 citeautoscript,
]{revtex4-2}

\usepackage{graphicx}
\usepackage{dcolumn}
\usepackage{bm}
\usepackage{xcolor}
\usepackage{hyperref}
\hypersetup{
    colorlinks=true,
    linkcolor=blue,
    citecolor=blue,
    filecolor=blue,      
    urlcolor=blue,
    }

\newcommand{\FS}[1]{\left\langle #1 \right\rangle_{\scriptscriptstyle\mathrm{FS}}}
 
\newcommand{\kb}{k_\mathrm{B}}
\newcommand{\tc}{T_\mathrm{c}}

\newcommand{\RR}{\mathbf{R}}
\newcommand{\vF}{\mathbf{v}_\mathrm{F}}
\newcommand{\pF}{\mathbf{p}_\mathrm{F}}

\newcommand{\Tr}{\mathrm{Tr}~}

\newcommand{\Dt}{\tilde{\Delta}}
\newcommand{\St}{\tilde{\Sigma}}

\begin{document}

\title{Spontaneous splitting of d-wave surface states:\\ Circulating currents or edge magnetization?}
\author{Kevin Marc Seja}
\thanks{These authors contributed equally.}
\author{Niclas Wall-Wennerdal}
\thanks{These authors contributed equally.}
\author{Tomas L\"ofwander}
\affiliation{Department of Microtechnology and Nanoscience - MC2,
Chalmers University of Technology,
SE-41296 G\"oteborg, Sweden}

\author{Mikael Fogelstr\"{o}m}
\affiliation{Department of Microtechnology and Nanoscience - MC2,
Chalmers University of Technology,
SE-41296 G\"oteborg, Sweden}
\affiliation{
Nordita, KTH Royal Institute of Technology and Stockholm University, Hannes Alfv\'{e}ns v\"{a}g 12, 10691 Stockholm, Sweden
}
\date{\today}

\begin{abstract}
Pair-breaking edges of $d$-wave superconductors feature Andreev bound states at the Fermi energy. Since these states are energetically highly unfavorable they are susceptible to effects that shift them to finite energy. We investigate the free energy of two different mechanisms: spontaneous phase gradients in the superconducting order parameter and surface ferromagnetism caused by Fermi liquid interaction effects. We find that the surface magnetization appears at lower temperatures than the spontaneous current flow of the phase-crystal state. The magnetic state can, however, be energetically favorable at lower temperatures for sufficiently strong Fermi liquid effects. As a result, first-order transitions between the two states are possible, suggesting a rich low-temperature phase diagram in $d$-wave superconductors.
\end{abstract}

\maketitle
\section{Introduction}
The rich physics of unconventional superconductors has been the subject of intense research for many years. 
A point of particular interest is the topologically protected surface states that have a large impact of the physics of such materials\cite{Hu1994, Lofwander2001May, Ryu2002Jul, Sato2011Jun}. 
In the case of d-wave superconductors, misaligned surfaces give rise to Andreev bound states at the Fermi energy as a result of scattering between lobes of the order parameter with different sign. These states carry a substantial spectral weight due to the large degeneracy with respect to the momentum parallel to the surface. As a result, they are energetically unfavorable and any mechanism that can move them away from the Fermi surface will reduce the ground state energy of the system\cite{Honerkamp1999May, lofwander_andreev_2001}. 
The Andreev states are experimentally observed as a zero bias tunneling conductance peak (ZBCP)\cite{Becherer1993Jun,Kashiwaya1994Dec, Kashiwaya1995Jan, kashiwaya_tunnelling_2000} and give rise to a paramagnetic Meissner effect\cite{Higashitani1997Sep}. 
The ZBCP peak has been observed to split into two separate ones in the presence of external magnetic fields but also spontaneously, i.e. without external field, at low temperatures\cite{Geerk1988Sep,Lesueur1992Feb, Covington1997Jul}. 
Over the years, different models for the underlying physics have been discussed, such as a subdominant $s$-wave order parameter at the surface that leads to a local breaking of time-reversal symmetry and shifts the Andreev bound states away from the Fermi energy\cite{Matsumoto1995Dec,Fogelstrom1997Jul}. 
As a competing mechanism, the possibility of ferromagnetic ordering at surfaces was suggested to appear due to electron-electron interaction\cite{Honerkamp1999May}. 
At zero temperature arbitrarily small interactions lead to spin splitting of surface states and a resulting magnetization, which was shown to be energetically favorable compared to subdominant s-wave order\cite{Potter2014Mar}. 
An underlying assumption in the scenarios above is translational invariance along the surface. This disallows another possible mechanism, the spontaneous development of spatially non-trivial phase gradients in the d-wave order parameter\cite{Hakansson2015Sep}. 
This phase-crystal state exhibits a periodic modulation of the phase, characterized by a wave vector $\mathbf{q}$, and associated current flow. 
The surfaces states are then Doppler shifted away from the Fermi energy, the resulting free energy gain at the surface exceeds the cost of the loop currents in the interior of the sample\cite{Holmvall2020Jan, Wennerdal2020Nov}. 
Strong correlations have been shown to stabilize the phase crystal, even in the presence of disorder\cite{Chakraborty2022Apr}. In Refs.\cite{Hakansson2015Sep, Holmvall2020Jan, Wennerdal2020Nov, Chakraborty2022Apr} spin degeneracy was assumed which neglects the possibility of magnetic ordering. It is thus an open question which of the two scenarios is going to be dominant, especially at experimentally relevant finite temperatures.
\\
In the present work, we use the quasiclassical theory of superconductivity and allow for both spontaneous phase gradients, as well as magnetic ordering at the surface. We consider a thin d-wave film at finite temperatures and for different magnetic interaction strengths in order to determine the state with minimal free energy. 
We find that the minimal-energy state is usually a phase-crystal state, with magnetic ordering dominating for large magnetic interaction strength and at low temperatures. As a result, there can be a crossover from one state to the other that appears as a first-order transition.
Our results thus give insight into the competition between the two different orders as well as fingerprints of the different phases.
\section{Theory} 
\subsection{Quasiclassical theory}
For our study, we use the quasiclassical theory of superconductivity in the Eilenberger form\cite{eilenberger_transformation_1968,Larkin1969,Eliashberg1971}. We only give a brief overview here, for details can be found in our earlier publications\cite{Seja2021Sep, Seja2022Oct} and other extensive literature\cite{Serene1983Dec, Kopnin2001,Eschrig2009Oct}. 
In the equilibrium situations considered here, all physical observables of interest can be calculated from the quasiclassical Green's function $\hat{g}^\mathrm{M}(\pF,\RR,\varepsilon_n)$ that depends on the momentum direction on the Fermi surface $\pF$, spatial coordinate $\RR$, and the Matsubara frequency $\varepsilon_n$. We obtain $\hat{g}^\mathrm{M}$ as a solution to the Eilenberger equation
\begin{align}
i \hbar &\vF \cdot \nabla \hat{g}^\mathrm{M}(\pF,\RR,\varepsilon_n) 
\nonumber
\\
&+ \bigl[i \varepsilon_n \hat{\tau}_3  - \hat{h}^\mathrm{M}(\pF,\RR,\varepsilon_n), \hat{g}^\mathrm{M}(\pF,\RR,\varepsilon_n) \bigr] = 0,
\label{eq:transportequation}
\end{align}
subject to the normalization condition
\begin{equation}
\hat{g}^\mathrm{M}(\pF,\RR,\varepsilon) \hat{g}^\mathrm{M}(\pF,\RR,\varepsilon) = -\pi^2.
\label{eq:normalizationCondition}
\end{equation}
In Eq.~\eqref{eq:transportequation}, a commutator between matrices $A$ and $B$ is denoted $[A,B]$ and $\hat{~}$ indicates a Nambu (particle-hole) space matrix such as the third Pauli matrix $\hat{\tau}_3$. The equation also contains the Fermi velocity $\vF$ and the self-energy matrix $\hat{h}^\mathrm{M}$ that we discuss in detail below. The Green's function $\hat{g}^\mathrm{M}$ in Eq.~\eqref{eq:transportequation} and Eq.~\eqref{eq:normalizationCondition} is a two-by-two matrix in particle hole space,
\begin{align}
\hat{g}^\mathrm{M}
=
\begin{pmatrix}
g^\mathrm{M} & f^\mathrm{M}
\\
\tilde{f}^\mathrm{M} & \tilde{g}^\mathrm{M}
\end{pmatrix},
\end{align}
with the quasiparticle Green's function on the diagonal and anomalous superconducting correlations on the off-diagonal. Each of these four elements is, in turn, a two-by-two matrix in spin space, for example
\begin{align} 
g^\mathrm{M} &= 
\begin{pmatrix}
    g_0\! + g_z & g_x\! -\! i g_y
    \\
    g_x\! +\! i g_y & g_0\! -\! g_z
\end{pmatrix}^\mathrm{M}
=
\begin{pmatrix}
    g_\uparrow & g_x \!- \!i g_y
    \\
    g_x\!+\!i g_y & g_\downarrow
\end{pmatrix}^\mathrm{M}.
\label{eq:GF_spinresolved}
\end{align}
The self-energy matrix $\hat{h}^\mathrm{M}$ has an identical structure in particle-hole and spin space. In this work we consider two selfenergy contributions,
\begin{align}
\hat{h}^\mathrm{M} = \hat{h}^\mathrm{M}_\mathrm{MF} + \hat{h}^\mathrm{M}_\mathrm{FL},
\end{align}
with a spin-singlet, mean-field order parameter
\begin{align}
\hat{h}^\mathrm{M}_\mathrm{MF}
= 
\begin{pmatrix}
0 & \Delta_s i \sigma_2\\
i \sigma_2 \tilde{\Delta}_s & 0 
\end{pmatrix},    
\end{align}
and a spin-dependent Fermi-liquid interaction
\begin{equation}
\hat{h}^\mathrm{M}_\mathrm{FL} = 
\begin{pmatrix}
\boldsymbol{\nu} \cdot \boldsymbol{\sigma} & 0
\\
0 & \boldsymbol{\nu} \cdot \boldsymbol{\sigma}^*
\end{pmatrix},
\end{equation}
Following Refs.~\cite{Serene1983Dec, Eschrig2001, Montiel2018Sep}
, the $i$-th element of $\boldsymbol{\nu}$ is
\begin{align}
\nu_i 
&\equiv A_{0,i} \kb T \sum\limits_{n} \FS{ \mathrm{Tr}_{\sigma_i} ~ \sigma_i \mathbf{g}  }.
\label{eq:FL_definition}
\end{align}
Here, $\FS{\dots}$ denotes an average over the Fermi surface, and $\mathrm{Tr}_\sigma$ is a trace over spin space. While $\nu_i$ has the dimension of energy, $A_{0,i} $ is a dimensionless scalar parameter that specifies the strength of the Fermi-liquid interaction along the spin axis $i$. It is related to the first spin-antisymmetric Landau parameter $F_0^a$ as 
\begin{align}
A_0 = \frac{F_0^a}{1 + F_0^a},
\label{eq:FL_parameter_relation}
\end{align}
and displays a ferromagnetic Stoner instability for $F_0^a \rightarrow -1$\cite{Coleman2015Nov}.
 For simplicity, we assume that a non-zero $A_{0,i}$ exists only along the spin-quantization axis which we label $z$. Generally, the interaction is ferromagnetic for negative values of $A_0$, so we replace $A_{0,z} \rightarrow -|A_0|$ and specify $|A_0|$ in the following\cite{Montiel2018Sep}. Assuming an antiferromagnetic interaction (positive $A_0$) results in a vanishing Fermi-liquid selfenergy in our system. In total we thus have a self-energy contribution
\begin{align}
\nu_z  =      
-2|A_{0} | \kb T  \sum\limits_{n} \FS{ g_z  }.
\end{align}
To solve Eq.~\eqref{eq:transportequation} we use  a Riccati parametrization for the Green's function $\hat{g}^M$\cite{Nagato1993, Schopohl1995, Schopohl1998, Shelankov2000Mar}, and a finite element method (FEM) that, compared to previous work\cite{Seja2022Oct}, was extended to systems with full spin structure. Details on this extended method can be found in the appendix of the present work.
To ensure that physical conservation laws are satisfied we solve for both the self-energies as well as the Green's function until self-consistency\cite{Sanchez-Canizares2001Jan, Seja2021Sep}.
Once such a self-consistent solution has been found, we calculate the difference in free energy from the normal state $\Delta\Omega_\mathrm{LW} = \Omega_S - \Omega_N$ using the Luttinger-Ward form of the free energy. Following Refs.~\cite{Luttinger1960Jun, Serene1983Dec,Thuneberg1984Apr, Vorontsov2003Aug, Ali2011Feb}, we arrive at
\begin{align}
~~~&\Delta\Omega_\mathrm{LW} = \Omega_S - \Omega_N
\nonumber
\\
&=
\frac{1}{2} \kb T \!
\sum\limits_{\varepsilon_n}  \Bigl( 
\int\limits_0^1  \FS{ \Tr \hat{h} \hat{g}_\lambda } \!\mathrm{d}\lambda
\!-\!\frac{1}{2}\FS{\Tr \hat{h}\hat{g}}
\Bigr).
\label{eq:LW_freeEnergy}
\end{align}
Here, $\hat{h}$ and $\hat{g}$ are the self-consistently determined selfenergies and Green's function, respectively. In contrast, $\hat{g}_\lambda$ is a solution of Eq.~\eqref{eq:transportequation} for scaled selfenergies, $\hat{\Sigma}  \rightarrow \lambda \hat{\Sigma}$, meaning a scaling of all self-energies from zero to the original value. Note that the boundary values of the Green's function have to be iteratively found for this scaled problem, while the self-energies are kept at the fixed scaling of the self-consistent solution.  
In the present case, we have a spin-singlet order parameter $\Delta_s$ and a diagonal self-energy that is proportional to $\sigma_z$, thus  
\begin{align}
\mathrm{Tr}~\hat{h}\hat{g} = 2\left( \nu_z g_z + \tilde{\nu}_z \tilde{g}_z - f_s \tilde{\Delta}_s - \tilde{f}_s \Delta_s \right),
\end{align}
and a similar expression for the term $\hat{h} \hat{g}_\lambda$. 
\\
Lastly, we calculate the spin-resolved density of states by solving Eq.~\eqref{eq:transportequation} for the retarded Green's function. To this end, we replace $i \varepsilon_n \rightarrow \varepsilon + i \eta$, in this work we use $\eta = 0.01 \kb \tc$ as a broadening parameter. From the retarded Green's function we obtain the Fermi-surface averaged density of states for spin component $\sigma={\uparrow, \downarrow}$ as
\begin{align}
\mathcal{N}_{\uparrow(\downarrow)}(\RR, \varepsilon) \equiv -\frac{\mathcal{N}_\mathrm{F}}{\pi}\FS{\mathrm{Im}~g_{\uparrow(\downarrow)}^R(\pF, \RR, \varepsilon)}
,
\label{eq:DoS_spinAndMomentum}
\end{align}	
where $g_{\uparrow(\downarrow)}$ is defined in Eq.~\eqref{eq:GF_spinresolved}. The full density of states is then simply
\begin{align}
\mathcal{N}(\RR, \varepsilon) = \mathcal{N}_{\uparrow}(\RR, \varepsilon) + \mathcal{N}_{\downarrow}(\RR, \varepsilon).
\end{align}
\subsection{Calculational strategy}
Our calculations proceed as follows. We initially assume a uniform order parameter and in-plane magnetic field that enters Eq.~\eqref{eq:transportequation} as a Zeeman-term
${\Sigma}_\mathrm{Zeeman} = \sigma_z h_\mathrm{Z} = \mu_\mathrm{B} B_\mathrm{ext}$.
This field serves as a seed for the surface magnetization by creating a spin splitting everywhere in the system. After ten selfconsistency iterations, we remove the external field by setting $B_\mathrm{ext} = 0$. Depending on the strength of the Fermi liquid interaction, $|A_0|$, the magnetization at the surface then either disappears, or remains and is present in the final self-consistent solution. For fixed $|A_0|$, we then calculate the free energy using Eq.~\eqref{eq:LW_freeEnergy}.
\subsection{Model}
We consider a thin-film $d$-wave superconductor in two different geometries. Firstly, a strip that is infinite in the $y$-direction but has a finite length $L = 40\xi_0$ in the $x$ direction, delimited by two fully reflective interfaces. This system is translationally invariant in the $y$-direction and quasi one-dimensional, hence we can solve Eq.~\eqref{eq:transportequation} along a one-dimensional line.
Secondly, we have a square with sides of length $L = 40\xi_0$ and fully reflective surfaces. In both systems, we assume specular scattering at the surfaces and a misalignment of $\alpha = \pi/4$ between the crystal axis and the respective surface normal\cite{Zhao2004Oct}. 
\begin{figure}[t]
    \centering
    \includegraphics{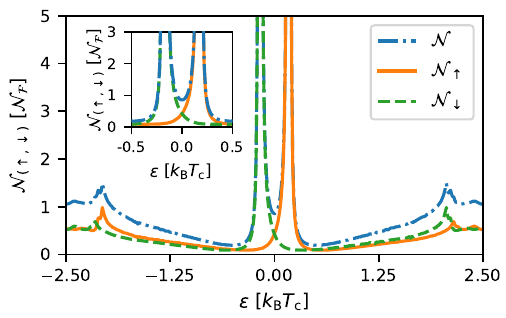}
    \caption{Fermi liquid interaction effects split the zero-energy Andreev bound states. The full density of states (dash-dotted blue) is split as a result of opposite energy shifts of up-spin (solid orange) and down-spin (dashed green) states. All quantities are shown directly at the surface of a translationally invariant strip with $|A_0| = 0.7$ at $T=0.15 \tc$. Inset: Enlarged view of the main plot around zero energy.}
    \label{fig:DOS_plot}
    \centering
    \includegraphics{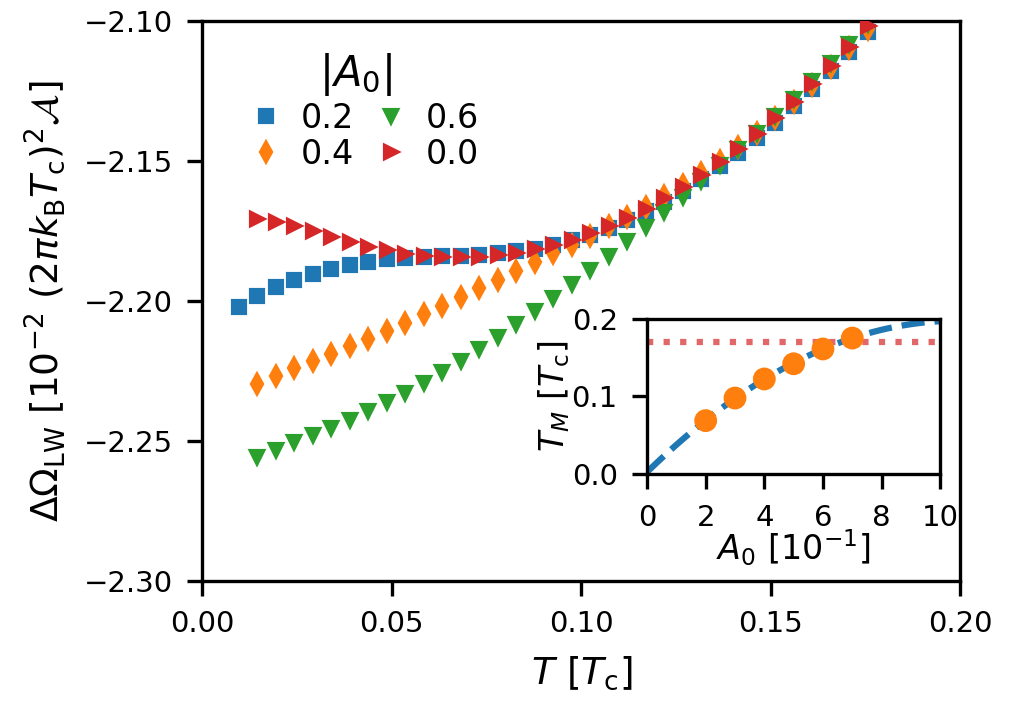}
    \caption{Free energy $\Delta \Omega_\mathrm{LW}$ as function of temperature $T$ in a infinite strip of length $L=20 \xi_0$ in the $x$-direction. In red triangles, the free energy in the absence of magnetic interaction, $|A_0|=0$, and for finite Fermi-liquid parameter $|A_0|$ as indicated in the legend. Inset: Dependence of $T_\mathrm{M}$, the transition temperature to the magnetic state, for various $A_0$ (orange dots) together with a quadratic fit (blue dashed line) and $T^*$, the temperature of the phase crystal in the absence of magnetic interaction (red-dotted line).}
    \label{fig:freeEnergy_1D}
\end{figure}
\section{Results}
\subsection{Translationally invariant strip (1D)}
In the translationally invariant system we cannot find a solution with spontaneous current flow. For a strip of width $L=40\xi_0$ we also do not find signs of the spontaneous symmetry-breaking phases that occur in narrow confined geometries\cite{Vorontsov2009Apr}. The only possible transition is then to a magnetized state. At a transition temperature $T_\mathrm{M}$ -- that depends on $|A_0|$ -- the surface Andreev bound states are shifted away from the Fermi energy. The spin-resolved density of states for up- and down-spin quasiparticles in Fig.~\ref{fig:DOS_plot} shows that the two peaks are spin polarized, this gives rise to a magnetization at the surface. The energy gained by this bound-state energy shift is larger then the cost of the induced magnetization that extends into the interior of the sample. 
The surface magnetization can be either of equal or opposite sign at the two edges, both configurations have equal free energy $\Delta \Omega_\mathrm{LW}$. The resulting free energy difference is shown in Fig.~\ref{fig:freeEnergy_1D} for several values of $|A_0|$. Both $\Delta \Omega_\mathrm{LW}$ and its derivative indicate a second-order phase transition to the magnetically ordered state at $T_\mathrm{M}$. The inset of Fig.~\ref{fig:freeEnergy_1D} shows the dependence of $T_\mathrm{M}$ on the Fermi-liquid parameter $|A_0|$. Clearly, for smaller $|A_0|$ the transition temperature $T_\mathrm{M}$ is lowered. This connects to the results of Ref.~\cite{Potter2014Mar} where, at zero temperature, an infinitesimally weak interaction is sufficient to create surface magnetization.

\begin{figure}[t]
    \centering
    \includegraphics{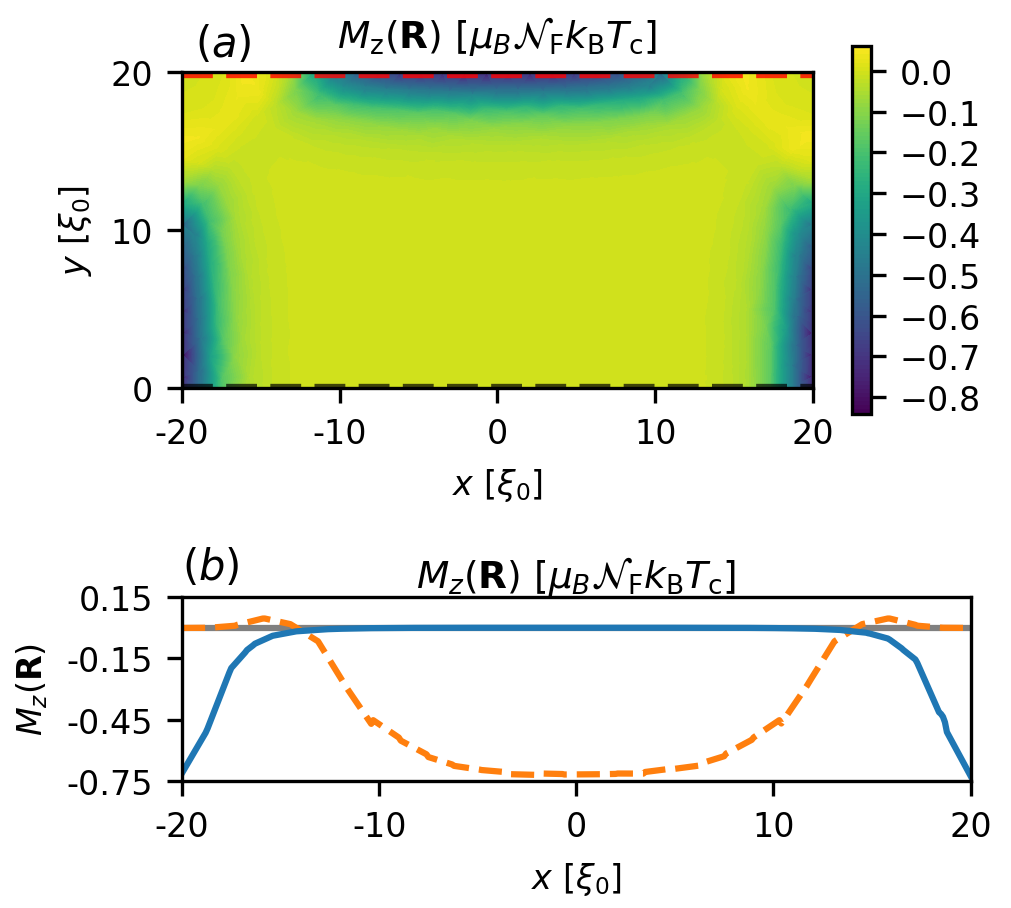}
    \caption{Surface magnetization for parameters $A_0 = 0.52$ and $T\approx 0.093 \tc$ in a square with an area $\mathcal{A} = (40 \xi_0)^2$. In (a), a filled contour plot of the magnetization $M_z(\mathbf{R})$, which is symmetric around $y=0$, in the upper half of the square. 
    In (b), in solid blue $M_z(\mathbf{R})$ along $y=0\xi_0$ -- seen as a black line in (a) -- showing the exponential decay into the bulk, and in dashed orange the magnetization along the surface at $y=-20\xi_0$, red line in (a), showing the decrease away from the center and small positive value close to the corners.}
    \label{fig:2d_magn}
\end{figure}
\subsection{Square geometry (2D)}
In a true square geometry we can find two non-trivial and distinct solutions as ground-state candidates. Firstly, a magnetically ordered state with magnetization on all edges, similar to the one found in the infinite strip. Secondly, we find the so-called phase crystal that features spontaneous phase gradients and currents along the edges. These currents typically flow in pairs of loops that have a radius of around five coherence lengths and carry, per loop-pair, counterflowing current.
We find that selfconsistent solutions are always either one or the other state and do not observe any mixed states. If the solution is of one or the other type depends on the temperature $T$ and Fermi-liquid interaction strength $|A_0|$. 
\begin{figure}[t!]
    \centering
    \includegraphics{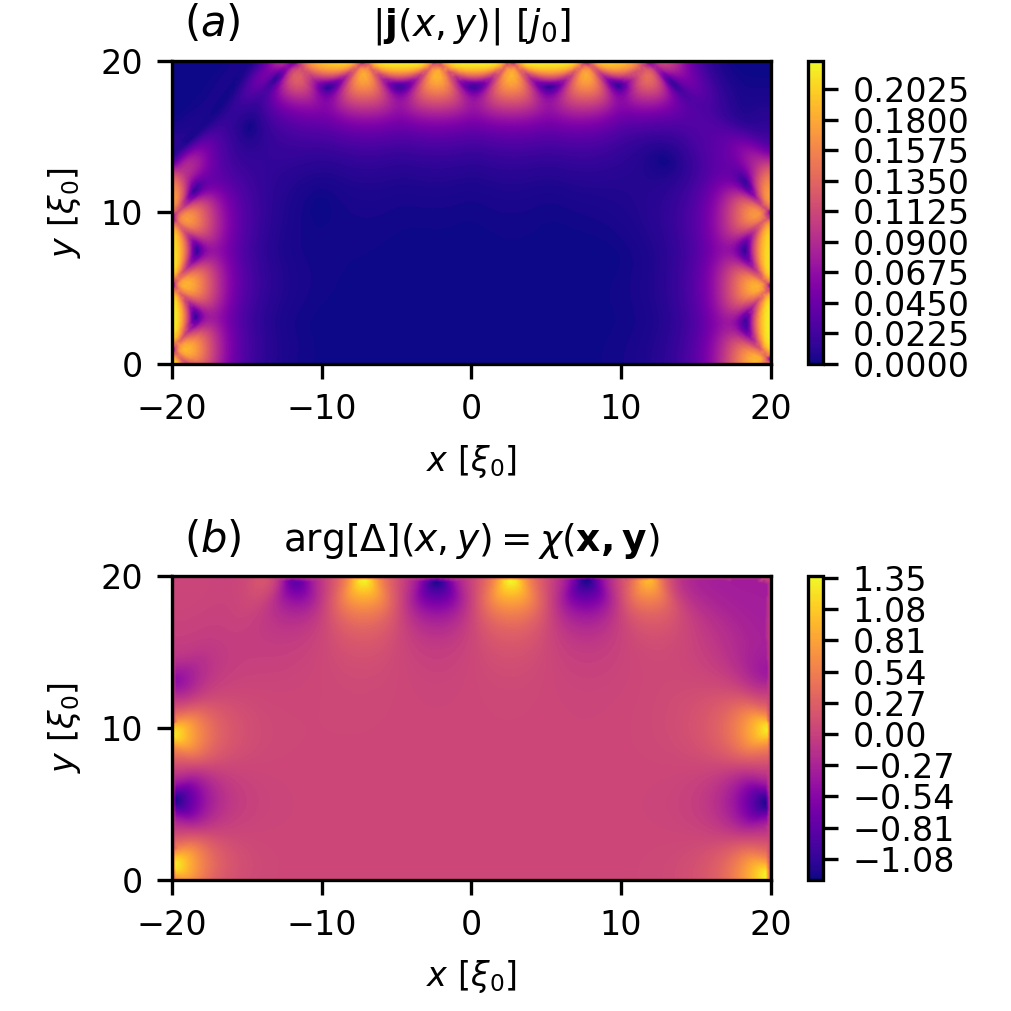}
    \caption{Example for the phase crystal state in a 2D square with area $\mathcal{A} = (40 \xi_0)^2$ for $T\approx 0.0884 \tc$. In (a), the norm of the current vector $\mathbf{j}(x,y)$ in units of $j_0 = \mathcal{N}_\mathrm{F} \vF \kb \tc$. In (b), the dimensionless phase of the order parameter $\chi(x,y) = \mathrm{arg}(\Delta)$. Both quantities are symmetric around $y=0$ and we show only the upper half of the square.}
    \label{fig:2d_currentloop}
    ~\\~\\
      \includegraphics{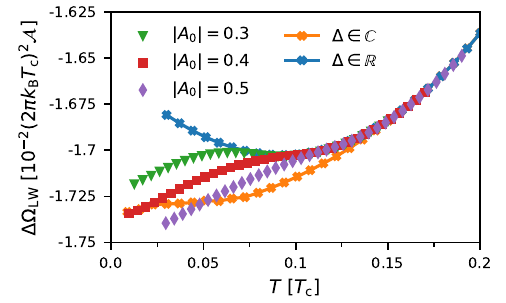}
    \caption{Free energy $\Delta \Omega_\mathrm{LW} $ for finite $A_0$ as indicated in the legend. Additionally, the case of vanishing $A_0$ for either a purely real (complex) order parameter marked with blue dots (orange crosses). At sufficiently low temperatures and large enough values of $|A_0|$ the free energy of the magnetic state is lower than that of the phase crystal.}
    \label{fig:2d_freeEnergy}
\end{figure}
We start by discussing the magnetic state, shown for one choice of parameters in Fig~\ref{fig:2d_magn}. As seen in Fig.~\ref{fig:2d_magn}(a), the magnetization is non-uniform along the edges and suppressed close to the grain corners due to interference effects. Fig.~\ref{fig:2d_magn}(b) shows that $M_z(\mathbf{R})$ is maximal in the center of each edge, and decays exponentially, on the scale of the superconducting coherence length, with distance from the surface. Similar to the infinite strip, we can find configurations with oppositely pointing magnetic fields at adjacent edges with no difference in free energy.
We now turn to the phase crystal state. The main characteristic are spontaneous currents that form loops along the edges, seen in Fig.~\ref{fig:2d_currentloop}(a). This spontaneous flow is the results of an oscillation of the order-parameter phase shown in Fig.~\ref{fig:2d_currentloop}(b). For details on the physics of this phase and how such currents can reduce the free energy, we refer to the existing literature \cite{Hakansson2015Sep, Holmvall2018Jun, Holmvall2020Jan}. The underlying solver package for the quasiclassical equations of motion in two dimensions, \textsc{SuperConga}, has been made publicly available\cite{Holmvall2023Mar}.
The phase crystal is also found in microscopic models and stabilized when including strong correlations\cite{Wennerdal2020Nov, Chakraborty2022Apr}. 

To compare the free energy of the two phases for different temperatures $T$ we consider two sets of parameters. First, we allow for a complex order parameter while Fermi-liquid effects are neglected, $A_0 = 0$, such that only a phase-crystal solution can appear below a critical temperature of $T^* \approx 0.17 \tc$\cite{Hakansson2015Sep}. In a separate set of calculations, we choose a finite $|A_0|$ and force the superconducting order parameter to be real. Depending on temperature, we then find either a magnetized state or a pure $d$-wave state without current loops. For the sake of comparison, we also calculated the free energy of a system with no magnetic interaction and real order parameter. In this case neither of the two mechanisms shifts the surface bound states and they stay at zero energy.
Fig.~\ref{fig:2d_freeEnergy} shows the free energy for these different sets of calculations.
For finite $|A_0|$, the surface magnetization lowers the free energy below the case of a purely real order parameter (blue line with crosses in Fig.~\ref{fig:2d_freeEnergy}). From a purely real order parameter, the state with surface magnetization is reached via a second-order phase transition at a temperature $T_\mathrm{M}$ that depends on $A_0$. This behavior is completely analogous to the infinite strip discussed earlier. 
In the two-dimensional square geometry, however, we can also find the phase crystal state. For a large temperature range this state with spontaneous current flow has even lower free energy, and is energetically favorable, compared to the state with surface magnetization. For sufficiently large $|A_0|$ and low enough temperature, the magnetic state can have a lower free energy than the phase crystal. In Fig.~\ref{fig:2d_freeEnergy} this is the case for $|A_0|= 0.4$ and $|A_0| = 0.5$.
In comparison, for smaller values such as $|A_0| = 0.3$ the phase crystal is always the state with lower free energy. Thus, the prevalence of e.g. the state with surface magnetization over the phase crystal depends strongly on the strength of Fermi liquid effects in a given material. 
For conventional superconductors, spin-polarized electron tunneling has been used to experimentally determine $|A_0|$, e.g. for dirty aluminum with a value of $|A_0| = 0.43 \pm 0.1$ \cite{Tedrow1984Apr, Alexander1985May, Meservey1994Mar}. An attempt to use similar experimental techniques on YBCO has been reported but gave inconclusive results\cite{Kashiwaya2004Sep}. What values to expect for typical $d$-wave materials is thus an open question to experiment. Recent microscopic calculations have predicted a large increase in the spin susceptibility close to pair-breaking surfaces compared to the bulk in such materials\cite{Matsubara2020Feb}. Since the spin susceptibility is to lowest order given by
$
\chi_\mathrm{s} \propto \mu_F^2 \mathcal{N}^*/( 1 + F_0^a),
$
this indicates a finite, negative Fermi-liquid parameter $F_0^a$\cite{Coleman2015Nov}. The values of Ref.~\cite{Matsubara2020Feb} correspond, through Eq.~\eqref{eq:FL_parameter_relation}, to $A_0\approx - 0.7$ which suggests that a low temperatures a dominance of the magnetic state over the phase crystal is likely. Experimentally, materials with strong enough Fermi liquid interactions should first display a second-order transition to a phase crystal state at $T \approx 0.17 T_c$. The state with surface magnetization then appears at a lower temperature  by a by a first-order transition.
An experimental fingerprint of the surface-magnetized states is the relatively uniform magnetization along pair-breaking edges compared to the phase crystal that features neighboring circular regions of oppositely-pointing magnetic fields generated by the current loop pattern\cite{Hakansson2015Sep}. The two phases should thus be experimentally distinguishable by the magnetic fields generated in the respective case.
\section{Discussion and Outlook}
We have studied the free energy of two distinct ground states in $d$-wave superconductors. We discussed a state with surface magnetization and another one with orbital currents circulating along the surfaces. In both cases, the free energy of the system is reduced by shifting the surface Andreev bound states to finite energy. At the same time, both the induced currents or magnetization cost energy in the interior of the sample\cite{Holmvall2020Jan}. The balance between the surface and interior free energy contributions determines the transition temperature to either of the two states. Both configurations are reached from a pure $d$-wave state via second-order phase transitions. For weak Fermi-liquid interaction, the transition to a magnetic state happens at a temperature $T_\mathrm{M}$ that is lower than the transition temperature of the phase crystal, $T^* \approx 0.17 \tc$. The magnetic state can, however, end up as the ground state for intermediate values of the Fermi liquid parameter and at lower temperatures. Which of the two states ends up being the ground state for given external parameters is determined by material properties, such as the strength of Fermi-liquid effects that renormalize the quasiparticle spectrum. The appearance of a $d\pm ip$-wave state at the surface has been discussed in Ref.~\cite{Matsubara2020Jun}, an analysis of this is beyond the scope of the present work. Additionally, Fermi liquid interaction beyond the lowest-order s-wave contribution considered can give rise to spin-orbit coupling and lead to additional splitting or broadening of the surface states.
Lastly, an external magnetic field applied in-plane on a thin film introduces a Zeeman splitting of the density of states that would make a surface magnetization energy more favorable while suppressing the phase crystal state. A detailed study of this biasing of the competition could thus alter the competition studied in this work, while also allowing for easier comparison to experiment.
\acknowledgements
We acknowledge the Swedish research council for financial support. The computations were enabled by resources provided by the National Academic Infrastructure for Supercomputing in Sweden (NAISS) at Tetralith partially funded by the Swedish Research Council through grant agreement no. 2022-06725. The results of this publication have been obtained with a FEM code that uses the open-source library \textsc{deal.ii}\cite{dealII95}.
\section*{Appendix}
\subsection*{Discontinuous Galerkin method for the quasiclassical theory with general spin structure}
We give here an overview of a finite-element based solution strategy to the general Eilenberger equation of quasiclassical theory with the full spin degrees of freedom. This extends a previously reported method for spin-degenerate system\cite{Seja2022Oct}. We focus here on the method for the imaginary-energy, or Matsubara, part of the Green's function that determine the equilibrium properties of a superconducting system. The generalization to nonequilibrum scenarios follows similar lines.
\\
The starting point is a parametrization of the Green's function $\hat{g}^\mathrm{M}$ in terms of coherence amplitudes $\gamma, \tilde{\gamma}$\cite{Schopohl1995, Schopohl1998}. For positive Matsubara frequencies, the Green's function can be written as
\begin{align}
\hat{g}^\mathrm{M} = 
-2\pi i
\!
\begin{pmatrix}
\mathcal{G}& \mathcal{F}
 \\
-\tilde{\mathcal{F}} & -\tilde{\mathcal{G}}
\end{pmatrix}^\mathrm{M}
+ i \pi \hat{\tau}_3
=
\begin{pmatrix}
g^\mathrm{M} & f^\mathrm{M}
\\
\tilde{f}^\mathrm{M} & \tilde{g}^\mathrm{M}.
\end{pmatrix}
\end{align}
where $\mathcal{G}^M \equiv ( 1 - \gamma^\mathrm{M} \tilde{\gamma}^\mathrm{M})^{-1}$ and $\mathcal{F}^\mathrm{M} \equiv \mathcal{G}^\mathrm{M}\gamma^\mathrm{M}$\cite{Eschrig2009Oct}.
In the following, we omit the superscript $\mathrm{M}$ to simplify notation. In the previously reported method for spin-degenerate systems, all elements of $\hat{g}$, and hence also the coherence amplitudes, are scalar quantities. 
For general spin structure, both the Green's function elements and the coherence amplitudes become instead two-by-two matrices in spin space. 
\\
We thus obtain a set of four coupled equations that have to be solved. The convention is to write e.g.
\begin{align}
g &= 
\begin{pmatrix}
    g_0 + g_z & g_x - i g_y
    \\
    g_x + i g_y & g_0 - g_z
\end{pmatrix},
\\
f &= 
\begin{pmatrix}
    f_0 + f_z & f_x - i f_y
    \\
    f_x + i f_y & f_0 - f_z
\end{pmatrix} i \sigma_2.
\end{align}
Similarly, we write for the coherence amplitude
\begin{align}
\gamma = 
\begin{pmatrix}
\gamma_0 + \gamma_z & \gamma_x - i \gamma_y
\\
\gamma_x + i \gamma_y & \gamma_0 - \gamma_z
\end{pmatrix}i \sigma_2
=
\begin{pmatrix}
\gamma_1 & \gamma_2
\\
\gamma_3 & \gamma_4
\end{pmatrix}i \sigma_2,
\label{eq:RiccatiAmplitudeIndexDefinition}
\end{align}
with an analogous form for $\tilde{\gamma}$. In the following, we will use the latter labeling of the elements of $\gamma$ in terms of numerical indices $k \in \{1,2,3,4\}$. Similarly, we then label the elements of the selfenergy matrices as
\begin{align}
\Delta =
\begin{pmatrix}
\Delta_1 & \Delta_2
\\
\Delta_3 & \Delta_4
\end{pmatrix}
i \sigma_2,
\qquad
\Sigma =
\begin{pmatrix}
\Sigma_1 & \Sigma_2
\\
\Sigma_3 & \Sigma_4
\end{pmatrix}.
\label{eq:SelfenergyIndexDefinition}
\end{align}
The projections onto singlet ($0$) or a given triplet ($x,y,z$) component can be obtained via the reverse linear transformation of Eq.~\eqref{eq:RiccatiAmplitudeIndexDefinition}, e.g. for the coherence amplitude
\begin{align}
\begin{pmatrix}
\gamma_0
&
\gamma_x 
\\
\gamma_y
&
\gamma_z
\end{pmatrix}
=
\frac{1}{2}
\begin{pmatrix}
\gamma_1 + \gamma_4
&
\gamma_2 + \gamma_3
\\
i(\gamma_2 - \gamma_3)
&
\gamma_1 - \gamma_4
\end{pmatrix}.
\end{align}
For general complex energies $\varepsilon$, the Riccati equation for the coherence amplitude $\gamma$ reads
\begin{align}
i \hbar \vF \cdot \nabla \gamma = \gamma \tilde{\Delta} \gamma - 2\varepsilon \gamma  + \Sigma \gamma - \gamma \tilde{\Sigma} - \Delta,
\label{eq:Riccati_equation_general}
\end{align}
with a symmetry-related equation for $\tilde{\gamma}$. Note that all objects in Eq.~\eqref{eq:Riccati_equation_general} expect for $\varepsilon$ are two-by-two spin matrices. Clearly, Eq.~\eqref{eq:Riccati_equation_general} is nonlinear in the unknown function $\gamma$ which we aim to solve through the iterative solution of a linearized problem using a finite element method. One possibility to get such an iterative sequence is to assume that the $n$-th iterative guess $\gamma^{(n)}$ is given as a solution to the linearized problem
\begin{align}
i \hbar \vF \cdot \nabla \gamma^{(n)} &= \gamma^{(n)} \tilde{\Delta} \gamma^{(n-1)} - 2\varepsilon \gamma^{(n)} 
\nonumber
\\
&+ \Sigma \gamma^{(n)} - \gamma^{(n)} \tilde{\Sigma} - \Delta.
\label{eq:Riccati_equation_linearized}
\end{align}
Given a starting guess $\gamma^{(0)}$ we then hope that the sequence $\gamma^{(n)}$ will convergence up to a desired accuracy in a reasonable amount of iterations.
To unburden the notation, we denote $\Gamma \equiv \gamma^{(n)}$ and $\gamma = \gamma^{(n-1)}$ in the following.
Using the labeling of Eqs.~(\ref{eq:RiccatiAmplitudeIndexDefinition}-\ref{eq:SelfenergyIndexDefinition})
and removing factors of $i \sigma_2$, Eq.~\eqref{eq:Riccati_equation_linearized} leads to an equation system of the form
\begin{widetext}
\begin{align}
i \hbar &\vF \cdot \nabla
\begin{pmatrix}
\Gamma_1 & \Gamma_2
\\
\Gamma_3 & \Gamma_4
\end{pmatrix}
=
\begin{pmatrix}
\Gamma_1 \left( - \Dt_4 \gamma_1 + \Dt_3 \gamma_3   \right) 
+
\Gamma_2 \left( \Dt_2 \gamma_1 - \Dt_1 \gamma_3  \right)
& 
\Gamma_1 \left( - \Dt_4 \gamma_2 + \Dt_3 \gamma_4  \right)
+
\Gamma_2 \left( \Dt_2 \gamma_2 - \Dt_1 \gamma_4  \right)
\\
\Gamma_4 \left( - \Dt_1 \gamma_3  + \Dt_2 \gamma_1  \right)
+
\Gamma_3 \left(  \Dt_3 \gamma_3 - \Dt_4 \gamma_1  \right)
&
\Gamma_4 \left( - \Dt_1 \gamma_4 + \Dt_2 \gamma_2 \right)
+
\Gamma_3 \left( \Dt_3 \gamma_4 - \Dt_4 \gamma_2   \right)
\end{pmatrix}
\nonumber
\\
& - 2  \varepsilon
\begin{pmatrix}
\Gamma_1 & \Gamma_2
\\
\Gamma_3 & \Gamma_4
\end{pmatrix}
+ 
\begin{pmatrix}
\Sigma_1 \Gamma_1 + \Sigma_2 \Gamma_3
& \Sigma_1 \Gamma_2 + \Sigma_2 \Gamma_4
\\
\Sigma_3 \Gamma_1 + \Sigma_4 \Gamma_3 
& \Sigma_4 \Gamma_4 + \Sigma_3 \Gamma_2 
\end{pmatrix}
-
\begin{pmatrix}
\Gamma_1 \St_4 -\St_2 \Gamma_2 & \Gamma_2 \St_1 - \Gamma_1 \St_3 
\\
\Gamma_3 \St_4 - \St_2 \Gamma_4 & \Gamma_4 \St_1 - \Gamma_3 \St_3
\end{pmatrix}
- 
\begin{pmatrix}
\Delta_1 
& \Delta_2
\\
\Delta_3 
& \Delta_4
\end{pmatrix}.
\label{eq:Riccati_picard_matrixequation}
\end{align}
\end{widetext}
Schematically, this system is equivalent to
\begin{align}
\label{eq:Riccati_schematic}
i \hbar \vF \cdot \nabla
\!\begin{pmatrix}
\Gamma_1 \\
\Gamma_2 \\
\Gamma_3 \\
\Gamma_4
\end{pmatrix}
= 
\begin{pmatrix}
r_1\left[\Gamma,\gamma, \Delta, \Sigma\right]
\\
r_2\left[\Gamma,\gamma, \Delta, \Sigma\right]
\\
r_3\left[\Gamma, \gamma, \Delta, \Sigma\right]
\\
r_4\left[\Gamma,\gamma, \Delta, \Sigma\right]
\end{pmatrix}
\!-
\!
\begin{pmatrix}
    \Delta_1
    \\
    \Delta_2
    \\
    \Delta_3
    \\
    \Delta_4
\end{pmatrix}.
\end{align}
As specified in Eq.~\eqref{eq:Riccati_picard_matrixequation} the four right-hand side functions $r_k$ -- with $k\in \{1,2,3,4\}$ -- depend on various elements of $\Gamma$, $\gamma$, $\Sigma$, and $\Delta$, which are in turn spatially dependent. In the following we only write out an explicit spatial dependence of $r_k(\RR)$. The ''driving term``, $\Delta_k$, is written out explicitly because it is the only one that is independent of $\Gamma_k$.
Note also that the differential operator $\vF \cdot \nabla$ acts on each element of the four-vector separately. 
By construction of Eq.~\eqref{eq:Riccati_equation_linearized} we have a \textit{linear} system of equations for the unknown functions $\Gamma_k$ in Eq.~\eqref{eq:Riccati_picard_matrixequation} and Eq.~\eqref{eq:Riccati_schematic}. 
\\
Performing a scalar product of both sides of the equation with a four vector of, currently unspecified, test functions
$\phi
= 
\Bigl( \phi_1(\RR)
    ,
    \phi_2(\RR)
    ,
    \phi_3(\RR)
    ,
    \phi_4(\RR)
\Bigr)^T,
$
and integrating over the domain $\Omega$ gives
\begin{align}
i \hbar 
\sum\limits_{k=1}^4
&\!\int\limits_\Omega
 \phi_k(\RR)  \left( \vF \cdot \nabla \Gamma_k(\RR) \right)
~\mathrm{d}\Omega
\nonumber
\\
&= 
\sum\limits_{k=1}^4
\int\limits_\Omega
 \phi_k(\RR) \left( r_k(\RR)-\Delta_k(\RR)\right)
\mathrm{d}\Omega.
\end{align}
The integration over the domain gets now split up into a sum of integrals over a set of cells $T_j$ that satisfy $\Omega = \cup_{T_j \in \mathcal{T}} T_j$, i.e., over a triangulation $\mathcal{T}$ of the domain $\Omega$. For the transport equation in Eq.~\eqref{eq:Riccati_schematic} it is crucial to use a so-called \textit{discontinuous Galerkin} method where neighboring cells have independent degrees of freedom associated with each geometric node. This means function values can be different in neighboring cells even at the - geometrically identical - shared cell corners\cite{Cockburn2000, Cockburn2003Nov}. 
\begin{figure}
    \centering
    \includegraphics[width=0.55\columnwidth]{{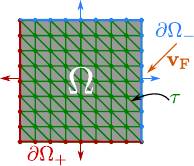}}
    \caption{ A domain $\Omega$ (grey) with its inflow (outflow) boundary $\partial \Omega_-$ ($\partial \Omega_+$) in marked in light blue (dark red) for the given transport direction $\vF$ (orange arrow). The collection of internal edges $\tau$ is marked in dark green. The mesh nodes of the underlying triangulation is marked by green circles. Small arrows on the domain boundary denote outward-pointing surface normals.}
    \label{fig:mesh_example}
\end{figure}
~\\
The splitting of the global integral into a sum of per-cell integrals gives
\begin{align}
i \hbar 
\sum\limits_{k=1}^4
&
\sum\limits_{T_j \in \mathcal{T}}\int\limits_{\Omega_j}
 \phi_k(\RR)
 \left( \vF \cdot \nabla \Gamma_k(\RR) \right)
~\mathrm{d}\Omega_j
\nonumber
\\
&=
\sum\limits_{k=1}^4
\sum\limits_{T_j \in \mathcal{T}} 
\int\limits_{\Omega_j}
\phi_k(\RR)\left( r_k(\RR)-\Delta_k(\RR)\right)
\mathrm{d}\Omega_j .
\end{align}
A partial integration of the left-hand side yields
\begin{align}
i \hbar 
\sum\limits_{k=1}^4
&\sum\limits_{T_j \in \mathcal{T}}
\Bigl[
\int\limits_{\partial \Omega_j}
 \phi_k(\RR)
  \Gamma_k(\RR) \vF \cdot  \mathbf{n}_j 
~\mathrm{d}s_j
\Bigr.
\nonumber
\\
&-
\Bigl.
\int\limits_{ \Omega_j}
  \Gamma_k(\RR) \vF \cdot  \nabla \phi_k(\RR) 
~\mathrm{d}\Omega_j
\Bigr]
\nonumber
\\
&=
\sum\limits_{i=1}^4
\sum\limits_{T_j \in \mathcal{T}} 
\int\limits_{\Omega_j}
\phi_k(\RR)\left( r_k(\RR)-\Delta_k(\RR)\right)
\mathrm{d}\Omega_j,
\label{eq:FEM_appendix_intermediate}
\end{align}
where the first integral is now over the boundary $\partial \Omega_j$ of a cell $T_j$ and contains the edge-dependent, outward-pointing normal vector $\mathbf{n}_j$. A given edge of such a cell will either be on the geometric boundary $\partial \Omega$ or one of the internal edges. We label the collection of such internal edges $\tau$. The geometric boundary $\partial \Omega$ is further split into an inflow boundary $\partial \Omega_-$ and an outflow boundary $\partial \Omega_+$, defined via
\begin{align}
   \partial \Omega_- &\equiv \left\{ {\RR} \in \partial \Omega ~|~\vF \cdot \mathbf{n}({\RR}) < 0 \right\} ,\\
    \partial \Omega_+ &\equiv \left\{ {\RR} \in \partial \Omega ~|~ \vF \cdot \mathbf{n}({\RR}) \geq 0 \right\}.
\end{align} 
The various sets are shown in Fig.~\ref{fig:mesh_example}. The sum over the cell-edge integration then consists of three different types of contributions. Firstly, integrals over edges on the inflow boundary where a boundary value $\Gamma_{k,B}$ has to be specified. We refer to the discussion in Ref.~\cite{Seja2022Oct} on how these boundary values are found since the procedure is identical to the spin-degenerate case. Secondly, we have integrals over the outflow boundary where the functions $\Gamma_k$ are unknown and determined in the later solution procedure. Lastly, Eq.~\eqref{eq:FEM_appendix_intermediate} features a sum over internal edges. Each edge is integrated over twice, once for each of the two cells that share a given edge. The sign of $\mathbf{n}_j \cdot \vF$ will be different for the two respective cells which leads to terms proportional to the difference of $\phi_k \Gamma_k$ in the two cells. We will just label two cells sharing an edge as cell 1 and cell 2. 
It has been shown\cite{Arnold2002, Brezzi2004Dec} that a numerically stabilized form of rewriting the integral contributions from internal edges in Eq.~\eqref{eq:FEM_appendix_intermediate} is

\begin{align}
\sum_{k=1}^4  &\sum\limits_{T_j \in \mathcal{T}} \int\limits_{ \partial\Omega_j}\!
\phi_k ~(\gamma_k \vF)\cdot \mathbf{n}_j~\mathrm{d}s_j
\nonumber
\\
&= \sum_{k=1}^4 \Biggl(
\sum\limits_{\tau_j \in \tau }  \int\limits_{\tau_j}
 \left\{\Gamma_k(\RR)  \vF \right\}_u \cdot  \left[\phi_k(\RR) \right]~  \mathrm{d}\tau_j
 \nonumber
 \\
 &
 +\!
 \sum\limits_{s_j \in \partial \Omega_+} \int\limits_{s_j}
 (\mathbf{n}_j \cdot \vF)~  \Gamma_k(\RR)   \phi_k(\RR)   ~\mathrm{d}s_j
 \nonumber
 \\
 &
 +\!\sum\limits_{ s_j\in \partial \Omega_-}  \int\limits_{s_j}
 (\mathbf{n}_j \cdot \vF)~  \Gamma_{k, \mathrm{B}}(\RR)  \phi_k(\RR)   ~\mathrm{d}s_j
 \Biggr)
 .
 \label{eq:boundaryIntegral_Form}
\end{align}
In Eq.~\eqref{eq:boundaryIntegral_Form} the first term on the right-hand side originates from the flow through internal edges, the other two terms originate from the domain boundary. The first term contains brackets with a subindex $u$, $\left\{ \dots \right\}_u$, that indicate the so-called upwind value
\begin{align}
\left\{\Gamma^R \vF \right\}_u 
\equiv
\left\{
	\begin{array}{ll}
		\Gamma_1^R \vF   & \mbox{if } \vF \cdot \mathbf{n}_1 > 0 \\
		\Gamma_2^R \vF & \mbox{if } \vF \cdot \mathbf{n}_1 < 0\\
		\left\{\Gamma^R \right\} \vF & \mbox{if } \vF \cdot \mathbf{n}_1 = 0
	\end{array}
\right. .
\label{eq:UpwindDefinition_main}
\end{align}
This definition and Eq.~\eqref{eq:boundaryIntegral_Form} contain the jump $[\dots]$ and average $\{\dots\}$ of a function along the edge shared by two cells. These bracket operators are defined for vectors $\mathbf{a}$ and scalars $\phi$ as
\begin{align}
\left[ \mathbf{a} \right]
\equiv 
\mathbf{a}_1 \cdot \mathbf{n}_1 + \mathbf{a}_2 \cdot \mathbf{n}_2,
\quad
\left[ \phi \right]
\equiv 
\phi_1 \mathbf{n}_1 + \phi_2 \mathbf{n}_2,
\label{eq:JumpDefinition}
\\
\left\{ \mathbf{a} \right\}
\equiv 
\frac{1}{2} \left( \mathbf{a}_1 + \mathbf{a}_2 \right),
\quad
\left\{ \phi \right\}
\equiv 
\frac{1}{2} \left( \phi_1 + \phi_2 \right).
\label{eq:AverageDefinition}
\end{align}
with the function value and the outward-pointing normal vector $\mathbf{n}_j$ in the respective cell labeled by an index (1,2).
Effectively, this entire stabilization procedure means that function values are i) specified on the inflow boundary, ii) propagated through internal edges along the given transport direction $\mathbf{v}_\mathrm{F}$, and iii) found on the outflow boundary as part of the solution step. This propagation is followed for positive Matsubara poles or the retarded components of the Green's function on the real axis. For advanced components, or negative Matsubara frequencies, the propagation directions swap. In the latter case boundary values are prescribed on the outflow boundary and values found on the inflow boundary while a downwind value, defined analogously to Eq.~\eqref{eq:UpwindDefinition_main}, propagates function values through internal edges.
In summary, this treatment mirrors the propagation of function values from a starting point to an end point along classical trajectories typically used in finite-difference approaches\cite{Eschrig2009Oct, Grein2013Aug, Holmvall2023Mar}. 
\\
 Lastly, the linear FEM weak form in Eq.~\eqref{eq:FEM_appendix_intermediate} should be written such that all terms containing the unknown functions $\Gamma_k$ are on the left-hand side while all terms that do not on the right-hand side of the equation. 
For our present problem, this means that the inflow-boundary term containing $\Gamma_{k,\mathrm{B}}$ in Eq.~\eqref{eq:boundaryIntegral_Form} needs to be moved to the right-hand side, while the term containing $r_k$ is moved to the left when combining Eq.~\eqref{eq:FEM_appendix_intermediate} and Eq.~\eqref{eq:boundaryIntegral_Form}. Doing so gives the translation of Eq.~\eqref{eq:Riccati_picard_matrixequation} into the corresponding weak form 
\begin{widetext}
    \begin{align}
i \hbar 
 &\sum_{k=1}^4 
\Biggl[
\sum\limits_{\tau_j \in \tau }  \int\limits_{\tau_j}
 \left\{\Gamma_k(\RR)  \vF \right\}_u \cdot  \left[\phi_k(\RR) \right]~  \mathrm{d}\tau_j
 \!+\!
 \sum\limits_{s_j \in \partial \Omega_+} \int\limits_{s_j}
 \mathbf{n}_j \cdot \vF~\Gamma_k(\RR)   \phi_k(\RR)   ~\mathrm{d}s_j
-
\sum\limits_{T_j \in \mathcal{T}} \int\limits_{ \Omega_j}
\Gamma_k(\RR)  \vF \cdot  \nabla \phi_k(\RR) 
~\mathrm{d}\Omega_j \Biggr]
 \nonumber
\\
&
- \sum_{k=1}^4 \sum\limits_{T_j \in \mathcal{T}} 
\int\limits_{\Omega_j}
\phi_k(\RR)  r_k(\RR) 
\mathrm{d}\Omega_j 
=
\sum\limits_{k=1}^4
\Biggl[
-
\sum\limits_{T_j \in \mathcal{T}} 
\int\limits_{\Omega_j}
\phi_k(\RR)  \Delta_k(\RR)
\mathrm{d}\Omega_j ~
-
\sum\limits_{ s_j\in \partial \Omega_-}  \int\limits_{s_j}
 \mathbf{n}_j \cdot \vF~   \Gamma_{k,B}(\RR)  \phi_k(\RR)   ~\mathrm{d}s_j\Biggr]~.
 \label{eq:weakForm_intermediate}
\end{align}
\end{widetext}
This weak form can now be treated with textbook methods in order to assemble and solve the corresponding matrix equation system\cite{Johnson2009Jan}. 
By solving the resulting system, we obtain a candidate for a new guess $\Gamma$ for the original non-linear system in Eq.~\eqref{eq:Riccati_equation_general} under the assumption of previous guess $\gamma = \gamma^{(n-1)}$. 
In some cases, directly taking the solution $\Gamma$ as the next iterative guess $\gamma^{(n)}$ is numerically unstable. We observe such instability in particular when solving the transport equations for real energies rather than purely imaginary Matsubara or Ozaki poles\cite{Ozaki2007Jan}. Given a solution $\Gamma_k$ of the linearized problem, one way to stabilize the iterative procedure is to update $\gamma_k^{(n)}$ via 
\begin{align}
    \gamma_k^{(n)} = \gamma_k^{(n-1)} + \alpha \left( \Gamma_k - \gamma_k^{(n-1)}\right).
\end{align}
Here, $\alpha \in (0,1]$ is a numerical parameter. We find that for general complex energies small values of $\gamma \lesssim 0.4$ are required which increases the number of iterative guesses. Adaptive methods that scale $\alpha$, e.g. based on the difference between $\Gamma$ and $\gamma$, can lead to faster convergence. In contrast, for purely imaginary poles it is stable to choose $\alpha=1$, i.e. directly assigning $\Gamma_k$ as the next guess for $\gamma_k^{(n)}$.
\bibliography{references}

\begin{thebibliography}{59}%
\makeatletter
\providecommand \@ifxundefined [1]{%
 \@ifx{#1\undefined}
}%
\providecommand \@ifnum [1]{%
 \ifnum #1\expandafter \@firstoftwo
 \else \expandafter \@secondoftwo
 \fi
}%
\providecommand \@ifx [1]{%
 \ifx #1\expandafter \@firstoftwo
 \else \expandafter \@secondoftwo
 \fi
}%
\providecommand \natexlab [1]{#1}%
\providecommand \enquote  [1]{``#1''}%
\providecommand \bibnamefont  [1]{#1}%
\providecommand \bibfnamefont [1]{#1}%
\providecommand \citenamefont [1]{#1}%
\providecommand \href@noop [0]{\@secondoftwo}%
\providecommand \href [0]{\begingroup \@sanitize@url \@href}%
\providecommand \@href[1]{\@@startlink{#1}\@@href}%
\providecommand \@@href[1]{\endgroup#1\@@endlink}%
\providecommand \@sanitize@url [0]{\catcode `\\12\catcode `\$12\catcode
  `\&12\catcode `\#12\catcode `\^12\catcode `\_12\catcode `\%12\relax}%
\providecommand \@@startlink[1]{}%
\providecommand \@@endlink[0]{}%
\providecommand \url  [0]{\begingroup\@sanitize@url \@url }%
\providecommand \@url [1]{\endgroup\@href {#1}{\urlprefix }}%
\providecommand \urlprefix  [0]{URL }%
\providecommand \Eprint [0]{\href }%
\providecommand \doibase [0]{https://doi.org/}%
\providecommand \selectlanguage [0]{\@gobble}%
\providecommand \bibinfo  [0]{\@secondoftwo}%
\providecommand \bibfield  [0]{\@secondoftwo}%
\providecommand \translation [1]{[#1]}%
\providecommand \BibitemOpen [0]{}%
\providecommand \bibitemStop [0]{}%
\providecommand \bibitemNoStop [0]{.\EOS\space}%
\providecommand \EOS [0]{\spacefactor3000\relax}%
\providecommand \BibitemShut  [1]{\csname bibitem#1\endcsname}%
\let\auto@bib@innerbib\@empty
\bibitem [{\citenamefont {Hu}(1994)}]{Hu1994}%
  \BibitemOpen
  \bibfield  {author} {\bibinfo {author} {\bibfnamefont {C.-R.}\ \bibnamefont
  {Hu}},\ }\bibfield  {title} {\bibinfo {title} {{Midgap surface states as a
  novel signature for
  ${\mathit{d}}_{\mathit{x}\mathit{a}}^{2}$-${\mathit{x}}_{\mathit{b}}^{2}$-wave
  superconductivity}},\ }\href {https://doi.org/10.1103/PhysRevLett.72.1526}
  {\bibfield  {journal} {\bibinfo  {journal} {Phys. Rev. Lett.}\ }\textbf
  {\bibinfo {volume} {72}},\ \bibinfo {pages} {1526} (\bibinfo {year}
  {1994})}\BibitemShut {NoStop}%
\bibitem [{\citenamefont {L{\ifmmode\ddot{o}\else\"{o}\fi}fwander}\ \emph
  {et~al.}(2001)\citenamefont {L{\ifmmode\ddot{o}\else\"{o}\fi}fwander},
  \citenamefont {Shumeiko},\ and\ \citenamefont {Wendin}}]{Lofwander2001May}%
  \BibitemOpen
  \bibfield  {author} {\bibinfo {author} {\bibfnamefont {T.}~\bibnamefont
  {L{\ifmmode\ddot{o}\else\"{o}\fi}fwander}}, \bibinfo {author} {\bibfnamefont
  {V.~S.}\ \bibnamefont {Shumeiko}},\ and\ \bibinfo {author} {\bibfnamefont
  {G.}~\bibnamefont {Wendin}},\ }\bibfield  {title} {\bibinfo {title} {{Andreev
  bound states in high-Tc superconducting}},\ }\href
  {https://doi.org/10.1088/0953-2048/14/5/201} {\bibfield  {journal} {\bibinfo
  {journal} {Supercond. Sci. Technol.}\ }\textbf {\bibinfo {volume} {14}},\
  \bibinfo {pages} {R53} (\bibinfo {year} {2001})}\BibitemShut {NoStop}%
\bibitem [{\citenamefont {Ryu}\ and\ \citenamefont
  {Hatsugai}(2002)}]{Ryu2002Jul}%
  \BibitemOpen
  \bibfield  {author} {\bibinfo {author} {\bibfnamefont {S.}~\bibnamefont
  {Ryu}}\ and\ \bibinfo {author} {\bibfnamefont {Y.}~\bibnamefont {Hatsugai}},\
  }\bibfield  {title} {\bibinfo {title} {{Topological Origin of Zero-Energy
  Edge States in Particle-Hole Symmetric Systems}},\ }\href
  {https://doi.org/10.1103/PhysRevLett.89.077002} {\bibfield  {journal}
  {\bibinfo  {journal} {Phys. Rev. Lett.}\ }\textbf {\bibinfo {volume} {89}},\
  \bibinfo {pages} {077002} (\bibinfo {year} {2002})}\BibitemShut {NoStop}%
\bibitem [{\citenamefont {Sato}\ \emph {et~al.}(2011)\citenamefont {Sato},
  \citenamefont {Tanaka}, \citenamefont {Yada},\ and\ \citenamefont
  {Yokoyama}}]{Sato2011Jun}%
  \BibitemOpen
  \bibfield  {author} {\bibinfo {author} {\bibfnamefont {M.}~\bibnamefont
  {Sato}}, \bibinfo {author} {\bibfnamefont {Y.}~\bibnamefont {Tanaka}},
  \bibinfo {author} {\bibfnamefont {K.}~\bibnamefont {Yada}},\ and\ \bibinfo
  {author} {\bibfnamefont {T.}~\bibnamefont {Yokoyama}},\ }\bibfield  {title}
  {\bibinfo {title} {{Topology of Andreev bound states with flat dispersion}},\
  }\href {https://doi.org/10.1103/PhysRevB.83.224511} {\bibfield  {journal}
  {\bibinfo  {journal} {Phys. Rev. B}\ }\textbf {\bibinfo {volume} {83}},\
  \bibinfo {pages} {224511} (\bibinfo {year} {2011})}\BibitemShut {NoStop}%
\bibitem [{\citenamefont {Honerkamp}\ and\ \citenamefont
  {Sigrist}(1999)}]{Honerkamp1999May}%
  \BibitemOpen
  \bibfield  {author} {\bibinfo {author} {\bibfnamefont {C.}~\bibnamefont
  {Honerkamp}}\ and\ \bibinfo {author} {\bibfnamefont {M.}~\bibnamefont
  {Sigrist}},\ }\bibfield  {title} {\bibinfo {title} {{Time-reversal symmetry
  breaking states at [110] surfaces of dx2{-}y2 superconductors}},\ }\href
  {https://doi.org/10.1016/S0921-4534(99)00106-9} {\bibfield  {journal}
  {\bibinfo  {journal} {Physica C}\ }\textbf {\bibinfo {volume} {317-318}},\
  \bibinfo {pages} {489} (\bibinfo {year} {1999})}\BibitemShut {NoStop}%
\bibitem [{\citenamefont {Löfwander}\ \emph {et~al.}(2001)\citenamefont
  {Löfwander}, \citenamefont {Shumeiko},\ and\ \citenamefont
  {Wendin}}]{lofwander_andreev_2001}%
  \BibitemOpen
  \bibfield  {author} {\bibinfo {author} {\bibfnamefont {T.}~\bibnamefont
  {Löfwander}}, \bibinfo {author} {\bibfnamefont {V.~S.}\ \bibnamefont
  {Shumeiko}},\ and\ \bibinfo {author} {\bibfnamefont {G.}~\bibnamefont
  {Wendin}},\ }\bibfield  {title} {\bibinfo {title} {{Andreev} bound states in
  high-{$T_\mathrm{c}$} superconducting junctions},\ }\href
  {https://doi.org/10.1088/0953-2048/14/5/201} {\bibfield  {journal} {\bibinfo
  {journal} {Supercond. Sci. Technol.}\ }\textbf {\bibinfo {volume} {14}},\
  \bibinfo {pages} {R53} (\bibinfo {year} {2001})}\BibitemShut {NoStop}%
\bibitem [{\citenamefont {Becherer}\ \emph {et~al.}(1993)\citenamefont
  {Becherer}, \citenamefont {St{\ifmmode\ddot{o}\else\"{o}\fi}lzel},
  \citenamefont {Adrian},\ and\ \citenamefont {Adrian}}]{Becherer1993Jun}%
  \BibitemOpen
  \bibfield  {author} {\bibinfo {author} {\bibfnamefont {{\relax
  T}.}~\bibnamefont {Becherer}}, \bibinfo {author} {\bibfnamefont
  {C.}~\bibnamefont {St{\ifmmode\ddot{o}\else\"{o}\fi}lzel}}, \bibinfo {author}
  {\bibfnamefont {G.}~\bibnamefont {Adrian}},\ and\ \bibinfo {author}
  {\bibfnamefont {H.}~\bibnamefont {Adrian}},\ }\bibfield  {title} {\bibinfo
  {title} {{Normal electron tunneling in ramp-type
  ${\mathrm{YBa}}_{2}$${\mathrm{Cu}}_{3}$${\mathrm{O}}_{7}$/${\mathrm{PrBa}}_{2}$${\mathrm{Cu}}_{3}$${\mathrm{O}}_{7}$/${\mathrm{YBa}}_{2}$${\mathrm{Cu}}_{3}$${\mathrm{O}}_{7}$
  junctions prepared by laser ablation}},\ }\href
  {https://doi.org/10.1103/PhysRevB.47.14650} {\bibfield  {journal} {\bibinfo
  {journal} {Phys. Rev. B}\ }\textbf {\bibinfo {volume} {47}},\ \bibinfo
  {pages} {14650} (\bibinfo {year} {1993})}\BibitemShut {NoStop}%
\bibitem [{\citenamefont {Kashiwaya}\ \emph {et~al.}(1994)\citenamefont
  {Kashiwaya}, \citenamefont {Tanaka}, \citenamefont {Koyanagi}, \citenamefont
  {Takashima},\ and\ \citenamefont {Kajimura}}]{Kashiwaya1994Dec}%
  \BibitemOpen
  \bibfield  {author} {\bibinfo {author} {\bibfnamefont {S.}~\bibnamefont
  {Kashiwaya}}, \bibinfo {author} {\bibfnamefont {Y.}~\bibnamefont {Tanaka}},
  \bibinfo {author} {\bibfnamefont {M.}~\bibnamefont {Koyanagi}}, \bibinfo
  {author} {\bibfnamefont {H.}~\bibnamefont {Takashima}},\ and\ \bibinfo
  {author} {\bibfnamefont {K.}~\bibnamefont {Kajimura}},\ }\bibfield  {title}
  {\bibinfo {title} {{Evidence for d-wave symmetry in high-Tc superconductors
  based on tunneling theory and STM experiment}},\ }\href
  {https://doi.org/10.1016/0921-4534(94)92177-6} {\bibfield  {journal}
  {\bibinfo  {journal} {Physica C}\ }\textbf {\bibinfo {volume} {235-240}},\
  \bibinfo {pages} {1911} (\bibinfo {year} {1994})}\BibitemShut {NoStop}%
\bibitem [{\citenamefont {Kashiwaya}\ \emph {et~al.}(1995)\citenamefont
  {Kashiwaya}, \citenamefont {Tanaka}, \citenamefont {Koyanagi}, \citenamefont
  {Takashima},\ and\ \citenamefont {Kajimura}}]{Kashiwaya1995Jan}%
  \BibitemOpen
  \bibfield  {author} {\bibinfo {author} {\bibfnamefont {S.}~\bibnamefont
  {Kashiwaya}}, \bibinfo {author} {\bibfnamefont {Y.}~\bibnamefont {Tanaka}},
  \bibinfo {author} {\bibfnamefont {M.}~\bibnamefont {Koyanagi}}, \bibinfo
  {author} {\bibfnamefont {H.}~\bibnamefont {Takashima}},\ and\ \bibinfo
  {author} {\bibfnamefont {K.}~\bibnamefont {Kajimura}},\ }\bibfield  {title}
  {\bibinfo {title} {{Origin of zero-bias conductance peaks in
  high-${\mathit{T}}_{\mathit{c}}$ superconductors}},\ }\href
  {https://doi.org/10.1103/PhysRevB.51.1350} {\bibfield  {journal} {\bibinfo
  {journal} {Phys. Rev. B}\ }\textbf {\bibinfo {volume} {51}},\ \bibinfo
  {pages} {1350} (\bibinfo {year} {1995})}\BibitemShut {NoStop}%
\bibitem [{\citenamefont {Kashiwaya}\ and\ \citenamefont
  {Tanaka}(2000)}]{kashiwaya_tunnelling_2000}%
  \BibitemOpen
  \bibfield  {author} {\bibinfo {author} {\bibfnamefont {S.}~\bibnamefont
  {Kashiwaya}}\ and\ \bibinfo {author} {\bibfnamefont {Y.}~\bibnamefont
  {Tanaka}},\ }\bibfield  {title} {\bibinfo {title} {Tunnelling effects on
  surface bound states in unconventional superconductors},\ }\href
  {https://doi.org/10.1088/0034-4885/63/10/202} {\bibfield  {journal} {\bibinfo
   {journal} {Rep. Prog. Phys.}\ }\textbf {\bibinfo {volume} {63}},\ \bibinfo
  {pages} {1641} (\bibinfo {year} {2000})}\BibitemShut {NoStop}%
\bibitem [{\citenamefont {Higashitani}(1997)}]{Higashitani1997Sep}%
  \BibitemOpen
  \bibfield  {author} {\bibinfo {author} {\bibfnamefont {S.}~\bibnamefont
  {Higashitani}},\ }\bibfield  {title} {\bibinfo {title} {{Mechanism of
  Paramagnetic Meissner Effect in High-Temperature Superconductors}},\ }\href
  {https://doi.org/10.1143/JPSJ.66.2556} {\bibfield  {journal} {\bibinfo
  {journal} {J. Phys. Soc. Jpn.}\ }\textbf {\bibinfo {volume} {66}},\ \bibinfo
  {pages} {2556} (\bibinfo {year} {1997})}\BibitemShut {NoStop}%
\bibitem [{\citenamefont {Geerk}\ \emph {et~al.}(1988)\citenamefont {Geerk},
  \citenamefont {Xi},\ and\ \citenamefont {Linker}}]{Geerk1988Sep}%
  \BibitemOpen
  \bibfield  {author} {\bibinfo {author} {\bibfnamefont {J.}~\bibnamefont
  {Geerk}}, \bibinfo {author} {\bibfnamefont {X.~X.}\ \bibnamefont {Xi}},\ and\
  \bibinfo {author} {\bibfnamefont {G.}~\bibnamefont {Linker}},\ }\bibfield
  {title} {\bibinfo {title} {{Electron tunneling into thin films of
  Y1Ba2Cu3O7}},\ }\href {https://doi.org/10.1007/BF01314271} {\bibfield
  {journal} {\bibinfo  {journal} {Z. Phys. B: Condens. Matter}\ }\textbf
  {\bibinfo {volume} {73}},\ \bibinfo {pages} {329} (\bibinfo {year}
  {1988})}\BibitemShut {NoStop}%
\bibitem [{\citenamefont {Lesueur}\ \emph {et~al.}(1992)\citenamefont
  {Lesueur}, \citenamefont {Greene}, \citenamefont {Feldmann},\ and\
  \citenamefont {Inam}}]{Lesueur1992Feb}%
  \BibitemOpen
  \bibfield  {author} {\bibinfo {author} {\bibfnamefont {J.}~\bibnamefont
  {Lesueur}}, \bibinfo {author} {\bibfnamefont {L.~H.}\ \bibnamefont {Greene}},
  \bibinfo {author} {\bibfnamefont {W.~L.}\ \bibnamefont {Feldmann}},\ and\
  \bibinfo {author} {\bibfnamefont {A.}~\bibnamefont {Inam}},\ }\bibfield
  {title} {\bibinfo {title} {{Zero bias anomalies in YBa2Cu3O7 tunnel
  junctions}},\ }\href {https://doi.org/10.1016/0921-4534(92)90926-4}
  {\bibfield  {journal} {\bibinfo  {journal} {Physica C}\ }\textbf {\bibinfo
  {volume} {191}},\ \bibinfo {pages} {325} (\bibinfo {year}
  {1992})}\BibitemShut {NoStop}%
\bibitem [{\citenamefont {Covington}\ \emph {et~al.}(1997)\citenamefont
  {Covington}, \citenamefont {Aprili}, \citenamefont {Paraoanu}, \citenamefont
  {Greene}, \citenamefont {Xu}, \citenamefont {Zhu},\ and\ \citenamefont
  {Mirkin}}]{Covington1997Jul}%
  \BibitemOpen
  \bibfield  {author} {\bibinfo {author} {\bibfnamefont {M.}~\bibnamefont
  {Covington}}, \bibinfo {author} {\bibfnamefont {M.}~\bibnamefont {Aprili}},
  \bibinfo {author} {\bibfnamefont {E.}~\bibnamefont {Paraoanu}}, \bibinfo
  {author} {\bibfnamefont {L.~H.}\ \bibnamefont {Greene}}, \bibinfo {author}
  {\bibfnamefont {F.}~\bibnamefont {Xu}}, \bibinfo {author} {\bibfnamefont
  {J.}~\bibnamefont {Zhu}},\ and\ \bibinfo {author} {\bibfnamefont {C.~A.}\
  \bibnamefont {Mirkin}},\ }\bibfield  {title} {\bibinfo {title} {{Observation
  of Surface-Induced Broken Time-Reversal Symmetry in
  ${\mathrm{YBa}}_{2}{\mathrm{Cu}}_{3}{O}_{7}$ Tunnel Junctions}},\ }\href
  {https://doi.org/10.1103/PhysRevLett.79.277} {\bibfield  {journal} {\bibinfo
  {journal} {Phys. Rev. Lett.}\ }\textbf {\bibinfo {volume} {79}},\ \bibinfo
  {pages} {277} (\bibinfo {year} {1997})}\BibitemShut {NoStop}%
\bibitem [{\citenamefont {Matsumoto}\ and\ \citenamefont
  {Shiba}(1995)}]{Matsumoto1995Dec}%
  \BibitemOpen
  \bibfield  {author} {\bibinfo {author} {\bibfnamefont {M.}~\bibnamefont
  {Matsumoto}}\ and\ \bibinfo {author} {\bibfnamefont {H.}~\bibnamefont
  {Shiba}},\ }\bibfield  {title} {\bibinfo {title} {{Coexistence of Different
  Symmetry Order Parameters near a Surface in d-Wave Superconductors II}},\
  }\href {https://doi.org/10.1143/JPSJ.64.4867} {\bibfield  {journal} {\bibinfo
   {journal} {J. Phys. Soc. Jpn.}\ }\textbf {\bibinfo {volume} {64}},\ \bibinfo
  {pages} {4867} (\bibinfo {year} {1995})}\BibitemShut {NoStop}%
\bibitem [{\citenamefont {Fogelstr{\ifmmode\ddot{o}\else\"{o}\fi}m}\ \emph
  {et~al.}(1997)\citenamefont {Fogelstr{\ifmmode\ddot{o}\else\"{o}\fi}m},
  \citenamefont {Rainer},\ and\ \citenamefont {Sauls}}]{Fogelstrom1997Jul}%
  \BibitemOpen
  \bibfield  {author} {\bibinfo {author} {\bibfnamefont {M.}~\bibnamefont
  {Fogelstr{\ifmmode\ddot{o}\else\"{o}\fi}m}}, \bibinfo {author} {\bibfnamefont
  {D.}~\bibnamefont {Rainer}},\ and\ \bibinfo {author} {\bibfnamefont {J.~A.}\
  \bibnamefont {Sauls}},\ }\bibfield  {title} {\bibinfo {title} {{Tunneling
  into Current-Carrying Surface States of High- ${\mathit{T}}_{\mathit{c}}$
  Superconductors}},\ }\href {https://doi.org/10.1103/PhysRevLett.79.281}
  {\bibfield  {journal} {\bibinfo  {journal} {Phys. Rev. Lett.}\ }\textbf
  {\bibinfo {volume} {79}},\ \bibinfo {pages} {281} (\bibinfo {year}
  {1997})}\BibitemShut {NoStop}%
\bibitem [{\citenamefont {Potter}\ and\ \citenamefont
  {Lee}(2014)}]{Potter2014Mar}%
  \BibitemOpen
  \bibfield  {author} {\bibinfo {author} {\bibfnamefont {A.~C.}\ \bibnamefont
  {Potter}}\ and\ \bibinfo {author} {\bibfnamefont {P.~A.}\ \bibnamefont
  {Lee}},\ }\bibfield  {title} {\bibinfo {title} {{Edge Ferromagnetism from
  Majorana Flat Bands: Application to Split Tunneling-Conductance Peaks in
  High-${T}_{c}$ Cuprate Superconductors}},\ }\href
  {https://doi.org/10.1103/PhysRevLett.112.117002} {\bibfield  {journal}
  {\bibinfo  {journal} {Phys. Rev. Lett.}\ }\textbf {\bibinfo {volume} {112}},\
  \bibinfo {pages} {117002} (\bibinfo {year} {2014})}\BibitemShut {NoStop}%
\bibitem [{\citenamefont {H{\aa}kansson}\ \emph {et~al.}(2015)\citenamefont
  {H{\aa}kansson}, \citenamefont {L{\ifmmode\ddot{o}\else\"{o}\fi}fwander},\
  and\ \citenamefont
  {Fogelstr{\ifmmode\ddot{o}\else\"{o}\fi}m}}]{Hakansson2015Sep}%
  \BibitemOpen
  \bibfield  {author} {\bibinfo {author} {\bibfnamefont {M.}~\bibnamefont
  {H{\aa}kansson}}, \bibinfo {author} {\bibfnamefont {T.}~\bibnamefont
  {L{\ifmmode\ddot{o}\else\"{o}\fi}fwander}},\ and\ \bibinfo {author}
  {\bibfnamefont {M.}~\bibnamefont
  {Fogelstr{\ifmmode\ddot{o}\else\"{o}\fi}m}},\ }\bibfield  {title} {\bibinfo
  {title} {{Spontaneously broken time-reversal symmetry in high-temperature
  superconductors - Nature Physics}},\ }\href
  {https://doi.org/10.1038/nphys3383} {\bibfield  {journal} {\bibinfo
  {journal} {Nat. Phys.}\ }\textbf {\bibinfo {volume} {11}},\ \bibinfo {pages}
  {755} (\bibinfo {year} {2015})}\BibitemShut {NoStop}%
\bibitem [{\citenamefont {Holmvall}\ \emph {et~al.}(2020)\citenamefont
  {Holmvall}, \citenamefont {Fogelstr{\ifmmode\ddot{o}\else\"{o}\fi}m},
  \citenamefont {L{\ifmmode\ddot{o}\else\"{o}\fi}fwander},\ and\ \citenamefont
  {Vorontsov}}]{Holmvall2020Jan}%
  \BibitemOpen
  \bibfield  {author} {\bibinfo {author} {\bibfnamefont {P.}~\bibnamefont
  {Holmvall}}, \bibinfo {author} {\bibfnamefont {M.}~\bibnamefont
  {Fogelstr{\ifmmode\ddot{o}\else\"{o}\fi}m}}, \bibinfo {author} {\bibfnamefont
  {T.}~\bibnamefont {L{\ifmmode\ddot{o}\else\"{o}\fi}fwander}},\ and\ \bibinfo
  {author} {\bibfnamefont {A.~B.}\ \bibnamefont {Vorontsov}},\ }\bibfield
  {title} {\bibinfo {title} {{Phase crystals}},\ }\href
  {https://doi.org/10.1103/PhysRevResearch.2.013104} {\bibfield  {journal}
  {\bibinfo  {journal} {Phys. Rev. Res.}\ }\textbf {\bibinfo {volume} {2}},\
  \bibinfo {pages} {013104} (\bibinfo {year} {2020})}\BibitemShut {NoStop}%
\bibitem [{\citenamefont {Wennerdal}\ \emph {et~al.}(2020)\citenamefont
  {Wennerdal}, \citenamefont {Ask}, \citenamefont {Holmvall}, \citenamefont
  {L{\ifmmode\ddot{o}\else\"{o}\fi}fwander},\ and\ \citenamefont
  {Fogelstr{\ifmmode\ddot{o}\else\"{o}\fi}m}}]{Wennerdal2020Nov}%
  \BibitemOpen
  \bibfield  {author} {\bibinfo {author} {\bibfnamefont {N.~W.}\ \bibnamefont
  {Wennerdal}}, \bibinfo {author} {\bibfnamefont {A.}~\bibnamefont {Ask}},
  \bibinfo {author} {\bibfnamefont {P.}~\bibnamefont {Holmvall}}, \bibinfo
  {author} {\bibfnamefont {T.}~\bibnamefont
  {L{\ifmmode\ddot{o}\else\"{o}\fi}fwander}},\ and\ \bibinfo {author}
  {\bibfnamefont {M.}~\bibnamefont
  {Fogelstr{\ifmmode\ddot{o}\else\"{o}\fi}m}},\ }\bibfield  {title} {\bibinfo
  {title} {{Breaking time-reversal and translational symmetry at edges of
  $d$-wave superconductors: Microscopic theory and comparison with
  quasiclassical theory}},\ }\href
  {https://doi.org/10.1103/PhysRevResearch.2.043198} {\bibfield  {journal}
  {\bibinfo  {journal} {Phys. Rev. Res.}\ }\textbf {\bibinfo {volume} {2}},\
  \bibinfo {pages} {043198} (\bibinfo {year} {2020})}\BibitemShut {NoStop}%
\bibitem [{\citenamefont {Chakraborty}\ \emph {et~al.}(2022)\citenamefont
  {Chakraborty}, \citenamefont {L{\ifmmode\ddot{o}\else\"{o}\fi}fwander},
  \citenamefont {Fogelstr{\ifmmode\ddot{o}\else\"{o}\fi}m},\ and\ \citenamefont
  {Black-Schaffer}}]{Chakraborty2022Apr}%
  \BibitemOpen
  \bibfield  {author} {\bibinfo {author} {\bibfnamefont {D.}~\bibnamefont
  {Chakraborty}}, \bibinfo {author} {\bibfnamefont {T.}~\bibnamefont
  {L{\ifmmode\ddot{o}\else\"{o}\fi}fwander}}, \bibinfo {author} {\bibfnamefont
  {M.}~\bibnamefont {Fogelstr{\ifmmode\ddot{o}\else\"{o}\fi}m}},\ and\ \bibinfo
  {author} {\bibfnamefont {A.~M.}\ \bibnamefont {Black-Schaffer}},\ }\bibfield
  {title} {\bibinfo {title} {{Disorder-robust phase crystal in high-temperature
  superconductors stabilized by strong correlations}},\ }\href
  {https://doi.org/10.1038/s41535-022-00450-w} {\bibfield  {journal} {\bibinfo
  {journal} {npj Quantum Mater.}\ }\textbf {\bibinfo {volume} {7}},\ \bibinfo
  {pages} {1} (\bibinfo {year} {2022})}\BibitemShut {NoStop}%
\bibitem [{\citenamefont
  {Eilenberger}(1968)}]{eilenberger_transformation_1968}%
  \BibitemOpen
  \bibfield  {author} {\bibinfo {author} {\bibfnamefont {G.}~\bibnamefont
  {Eilenberger}},\ }\bibfield  {title} {\bibinfo {title} {Transformation of
  {Gorkov}'s equation for type {II} superconductors into transport-like
  equations},\ }\href {https://doi.org/10.1007/BF01379803} {\bibfield
  {journal} {\bibinfo  {journal} {Z. Physik}\ }\textbf {\bibinfo {volume}
  {214}},\ \bibinfo {pages} {195} (\bibinfo {year} {1968})}\BibitemShut
  {NoStop}%
\bibitem [{\citenamefont {Larkin}\ and\ \citenamefont
  {Ovchinnikov}(1968)}]{Larkin1969}%
  \BibitemOpen
  \bibfield  {author} {\bibinfo {author} {\bibfnamefont {A.}~\bibnamefont
  {Larkin}}\ and\ \bibinfo {author} {\bibfnamefont {Y.~N.}\ \bibnamefont
  {Ovchinnikov}},\ }\bibfield  {title} {\bibinfo {title} {{Quasiclassical
  Method in the Theory of Superconductivity}},\ }\href
  {http://jetp.ras.ru/cgi-bin/e/index/e/28/6/p1200?a=list} {\bibfield
  {journal} {\bibinfo  {journal} {Zh. Eksp. Teor. Fiz.}\ }\textbf {\bibinfo
  {volume} {55}},\ \bibinfo {pages} {2262} (\bibinfo {year} {1968})},\ \bibinfo
  {note} {[Sov. Phys. JETP {\bf 28}, 1200, 1969]}\BibitemShut {NoStop}%
\bibitem [{\citenamefont {Eliashberg}(1971)}]{Eliashberg1971}%
  \BibitemOpen
  \bibfield  {author} {\bibinfo {author} {\bibfnamefont {G.}~\bibnamefont
  {Eliashberg}},\ }\bibfield  {title} {\bibinfo {title} {Inelastic electron
  collisions and nonequilibrium stationary states in superconductors},\ }\href
  {http://jetp.ras.ru/cgi-bin/e/index/e/34/3/p668?a=list} {\bibfield  {journal}
  {\bibinfo  {journal} {Zh. Eksp. Teor. Fiz.}\ }\textbf {\bibinfo {volume}
  {61}},\ \bibinfo {pages} {1254} (\bibinfo {year} {1971})},\ \bibinfo {note}
  {[Sov. Phys. JETP {\bf 34}, 668 (1972)]}\BibitemShut {NoStop}%
\bibitem [{\citenamefont {Seja}\ and\ \citenamefont
  {L{\ifmmode\ddot{o}\else\"{o}\fi}fwander}(2021)}]{Seja2021Sep}%
  \BibitemOpen
  \bibfield  {author} {\bibinfo {author} {\bibfnamefont {K.~M.}\ \bibnamefont
  {Seja}}\ and\ \bibinfo {author} {\bibfnamefont {T.}~\bibnamefont
  {L{\ifmmode\ddot{o}\else\"{o}\fi}fwander}},\ }\bibfield  {title} {\bibinfo
  {title} {{Quasiclassical theory of charge transport across mesoscopic
  normal-metal--superconducting heterostructures with current conservation}},\
  }\href {https://doi.org/10.1103/PhysRevB.104.104502} {\bibfield  {journal}
  {\bibinfo  {journal} {Phys. Rev. B}\ }\textbf {\bibinfo {volume} {104}},\
  \bibinfo {pages} {104502} (\bibinfo {year} {2021})}\BibitemShut {NoStop}%
\bibitem [{\citenamefont {Seja}\ and\ \citenamefont
  {L{\ifmmode\ddot{o}\else\"{o}\fi}fwander}(2022)}]{Seja2022Oct}%
  \BibitemOpen
  \bibfield  {author} {\bibinfo {author} {\bibfnamefont {K.~M.}\ \bibnamefont
  {Seja}}\ and\ \bibinfo {author} {\bibfnamefont {T.}~\bibnamefont
  {L{\ifmmode\ddot{o}\else\"{o}\fi}fwander}},\ }\bibfield  {title} {\bibinfo
  {title} {{Finite element method for the quasiclassical theory of
  superconductivity}},\ }\href {https://doi.org/10.1103/PhysRevB.106.144511}
  {\bibfield  {journal} {\bibinfo  {journal} {Phys. Rev. B}\ }\textbf {\bibinfo
  {volume} {106}},\ \bibinfo {pages} {144511} (\bibinfo {year}
  {2022})}\BibitemShut {NoStop}%
\bibitem [{\citenamefont {Serene}\ and\ \citenamefont
  {Rainer}(1983)}]{Serene1983Dec}%
  \BibitemOpen
  \bibfield  {author} {\bibinfo {author} {\bibfnamefont {J.~W.}\ \bibnamefont
  {Serene}}\ and\ \bibinfo {author} {\bibfnamefont {D.}~\bibnamefont
  {Rainer}},\ }\bibfield  {title} {\bibinfo {title} {{The quasiclassical
  approach to superfluid 3He}},\ }\href
  {https://doi.org/10.1016/0370-1573(83)90051-0} {\bibfield  {journal}
  {\bibinfo  {journal} {Phys. Rep.}\ }\textbf {\bibinfo {volume} {101}},\
  \bibinfo {pages} {221} (\bibinfo {year} {1983})}\BibitemShut {NoStop}%
\bibitem [{\citenamefont {Kopnin}(2001)}]{Kopnin2001}%
  \BibitemOpen
  \bibfield  {author} {\bibinfo {author} {\bibfnamefont {N.}~\bibnamefont
  {Kopnin}},\ }\href {https://doi.org/10.1093/arclin/13.1.67} {\emph {\bibinfo
  {title} {{Theory of Nonequilibrium Superconductivity}}}}\ (\bibinfo
  {publisher} {Oxford University Press},\ \bibinfo {address} {Oxford, England,
  UK},\ \bibinfo {year} {2001})\BibitemShut {NoStop}%
\bibitem [{\citenamefont {Eschrig}(2009)}]{Eschrig2009Oct}%
  \BibitemOpen
  \bibfield  {author} {\bibinfo {author} {\bibfnamefont {M.}~\bibnamefont
  {Eschrig}},\ }\bibfield  {title} {\bibinfo {title} {{Scattering problem in
  nonequilibrium quasiclassical theory of metals and superconductors: General
  boundary conditions and applications}},\ }\href
  {https://doi.org/10.1103/PhysRevB.80.134511} {\bibfield  {journal} {\bibinfo
  {journal} {Phys. Rev. B}\ }\textbf {\bibinfo {volume} {80}},\ \bibinfo
  {pages} {134511} (\bibinfo {year} {2009})}\BibitemShut {NoStop}%
\bibitem [{\citenamefont {Eschrig}\ \emph {et~al.}(2001)\citenamefont
  {Eschrig}, \citenamefont {Sauls}, \citenamefont {Burkhardt},\ and\
  \citenamefont {Rainer}}]{Eschrig2001}%
  \BibitemOpen
  \bibfield  {author} {\bibinfo {author} {\bibfnamefont {M.}~\bibnamefont
  {Eschrig}}, \bibinfo {author} {\bibfnamefont {J.~A.}\ \bibnamefont {Sauls}},
  \bibinfo {author} {\bibfnamefont {H.}~\bibnamefont {Burkhardt}},\ and\
  \bibinfo {author} {\bibfnamefont {D.}~\bibnamefont {Rainer}},\ }\bibfield
  {title} {\bibinfo {title} {{Fermi Liquid Superconductivity}},\ }in\ \href
  {https://doi.org/10.1007/978-94-010-0758-0_21} {\emph {\bibinfo {booktitle}
  {{High-Tc Superconductors and Related Materials: Material Science,
  Fundamental Properties, and Some Future Electronic Applications}}}}\
  (\bibinfo  {publisher} {Springer},\ \bibinfo {address} {Dordrecht, The
  Netherlands},\ \bibinfo {year} {2001})\ pp.\ \bibinfo {pages}
  {413--446}\BibitemShut {NoStop}%
\bibitem [{\citenamefont {Montiel}\ and\ \citenamefont
  {Eschrig}(2018)}]{Montiel2018Sep}%
  \BibitemOpen
  \bibfield  {author} {\bibinfo {author} {\bibfnamefont {X.}~\bibnamefont
  {Montiel}}\ and\ \bibinfo {author} {\bibfnamefont {M.}~\bibnamefont
  {Eschrig}},\ }\bibfield  {title} {\bibinfo {title} {{Generation of pure
  superconducting spin current in magnetic heterostructures via nonlocally
  induced magnetism due to Landau Fermi liquid effects}},\ }\href
  {https://doi.org/10.1103/PhysRevB.98.104513} {\bibfield  {journal} {\bibinfo
  {journal} {Phys. Rev. B}\ }\textbf {\bibinfo {volume} {98}},\ \bibinfo
  {pages} {104513} (\bibinfo {year} {2018})}\BibitemShut {NoStop}%
\bibitem [{\citenamefont {Coleman}(2015)}]{Coleman2015Nov}%
  \BibitemOpen
  \bibfield  {author} {\bibinfo {author} {\bibfnamefont {P.}~\bibnamefont
  {Coleman}},\ }\href {https://doi.org/10.1017/CBO9781139020916} {\emph
  {\bibinfo {title} {{Introduction to Many-Body Physics}}}}\ (\bibinfo
  {publisher} {Cambridge University Press},\ \bibinfo {address} {Cambridge,
  England, UK},\ \bibinfo {year} {2015})\BibitemShut {NoStop}%
\bibitem [{\citenamefont {Nagato}\ \emph {et~al.}(1993)\citenamefont {Nagato},
  \citenamefont {Nagai},\ and\ \citenamefont {Hara}}]{Nagato1993}%
  \BibitemOpen
  \bibfield  {author} {\bibinfo {author} {\bibfnamefont {Y.}~\bibnamefont
  {Nagato}}, \bibinfo {author} {\bibfnamefont {K.}~\bibnamefont {Nagai}},\ and\
  \bibinfo {author} {\bibfnamefont {J.}~\bibnamefont {Hara}},\ }\bibfield
  {title} {\bibinfo {title} {{Theory of the Andreev reflection and the density
  of states in proximity contact normal-superconducting infinite
  double-layer}},\ }\href {https://doi.org/10.1007/BF00682280} {\bibfield
  {journal} {\bibinfo  {journal} {J. Low Temp. Phys.}\ }\textbf {\bibinfo
  {volume} {93}},\ \bibinfo {pages} {33} (\bibinfo {year} {1993})}\BibitemShut
  {NoStop}%
\bibitem [{\citenamefont {Schopohl}\ and\ \citenamefont
  {Maki}(1995)}]{Schopohl1995}%
  \BibitemOpen
  \bibfield  {author} {\bibinfo {author} {\bibfnamefont {N.}~\bibnamefont
  {Schopohl}}\ and\ \bibinfo {author} {\bibfnamefont {K.}~\bibnamefont
  {Maki}},\ }\bibfield  {title} {\bibinfo {title} {{Quasiparticle spectrum
  around a vortex line in a d-wave superconductor}},\ }\href
  {https://doi.org/10.1103/PhysRevB.52.490} {\bibfield  {journal} {\bibinfo
  {journal} {Phys. Rev. B}\ }\textbf {\bibinfo {volume} {52}},\ \bibinfo
  {pages} {490} (\bibinfo {year} {1995})}\BibitemShut {NoStop}%
\bibitem [{\citenamefont {Schopohl}(1998)}]{Schopohl1998}%
  \BibitemOpen
  \bibfield  {author} {\bibinfo {author} {\bibfnamefont {N.}~\bibnamefont
  {Schopohl}},\ }\bibfield  {title} {\bibinfo {title} {{Transformation of the
  Eilenberger Equations of Superconductivity to a Scalar Riccati Equation}},\
  }\href {https://arxiv.org/abs/cond-mat/9804064v1} {\bibfield  {journal}
  {\bibinfo  {journal} {arXiv}\ } (\bibinfo {year} {1998})},\ \Eprint
  {https://arxiv.org/abs/cond-mat/9804064} {cond-mat/9804064} \BibitemShut
  {NoStop}%
\bibitem [{\citenamefont {Shelankov}\ and\ \citenamefont
  {Ozana}(2000)}]{Shelankov2000Mar}%
  \BibitemOpen
  \bibfield  {author} {\bibinfo {author} {\bibfnamefont {A.}~\bibnamefont
  {Shelankov}}\ and\ \bibinfo {author} {\bibfnamefont {M.}~\bibnamefont
  {Ozana}},\ }\bibfield  {title} {\bibinfo {title} {{Quasiclassical theory of
  superconductivity: A multiple-interface geometry}},\ }\href
  {https://doi.org/10.1103/PhysRevB.61.7077} {\bibfield  {journal} {\bibinfo
  {journal} {Phys. Rev. B}\ }\textbf {\bibinfo {volume} {61}},\ \bibinfo
  {pages} {7077} (\bibinfo {year} {2000})}\BibitemShut {NoStop}%
\bibitem [{\citenamefont
  {S{\ifmmode\acute{a}\else\'{a}\fi}nchez-Ca{\ifmmode\tilde{n}\else\~{n}\fi}izares}\
  and\ \citenamefont {Sols}(2001)}]{Sanchez-Canizares2001Jan}%
  \BibitemOpen
  \bibfield  {author} {\bibinfo {author} {\bibfnamefont {J.}~\bibnamefont
  {S{\ifmmode\acute{a}\else\'{a}\fi}nchez-Ca{\ifmmode\tilde{n}\else\~{n}\fi}izares}}\
  and\ \bibinfo {author} {\bibfnamefont {F.}~\bibnamefont {Sols}},\ }\bibfield
  {title} {\bibinfo {title} {{Self-Consistent Theory of Transport in
  Quasi{\textendash}One-Dimensional Superconducting Wires}},\ }\href
  {https://doi.org/10.1023/A:1004896902269} {\bibfield  {journal} {\bibinfo
  {journal} {J. Low Temp. Phys.}\ }\textbf {\bibinfo {volume} {122}},\ \bibinfo
  {pages} {11} (\bibinfo {year} {2001})}\BibitemShut {NoStop}%
\bibitem [{\citenamefont {Luttinger}\ and\ \citenamefont
  {Ward}(1960)}]{Luttinger1960Jun}%
  \BibitemOpen
  \bibfield  {author} {\bibinfo {author} {\bibfnamefont {J.~M.}\ \bibnamefont
  {Luttinger}}\ and\ \bibinfo {author} {\bibfnamefont {J.~C.}\ \bibnamefont
  {Ward}},\ }\bibfield  {title} {\bibinfo {title} {{Ground-State Energy of a
  Many-Fermion System. II}},\ }\href {https://doi.org/10.1103/PhysRev.118.1417}
  {\bibfield  {journal} {\bibinfo  {journal} {Phys. Rev.}\ }\textbf {\bibinfo
  {volume} {118}},\ \bibinfo {pages} {1417} (\bibinfo {year}
  {1960})}\BibitemShut {NoStop}%
\bibitem [{\citenamefont {Thuneberg}\ \emph {et~al.}(1984)\citenamefont
  {Thuneberg}, \citenamefont {Kurkij{\ifmmode\ddot{a}\else\"{a}\fi}rvi},\ and\
  \citenamefont {Rainer}}]{Thuneberg1984Apr}%
  \BibitemOpen
  \bibfield  {author} {\bibinfo {author} {\bibfnamefont {E.~V.}\ \bibnamefont
  {Thuneberg}}, \bibinfo {author} {\bibfnamefont {J.}~\bibnamefont
  {Kurkij{\ifmmode\ddot{a}\else\"{a}\fi}rvi}},\ and\ \bibinfo {author}
  {\bibfnamefont {D.}~\bibnamefont {Rainer}},\ }\bibfield  {title} {\bibinfo
  {title} {{Elementary-flux-pinning potential in type-II superconductors}},\
  }\href {https://doi.org/10.1103/PhysRevB.29.3913} {\bibfield  {journal}
  {\bibinfo  {journal} {Phys. Rev. B}\ }\textbf {\bibinfo {volume} {29}},\
  \bibinfo {pages} {3913} (\bibinfo {year} {1984})}\BibitemShut {NoStop}%
\bibitem [{\citenamefont {Vorontsov}\ and\ \citenamefont
  {Sauls}(2003)}]{Vorontsov2003Aug}%
  \BibitemOpen
  \bibfield  {author} {\bibinfo {author} {\bibfnamefont {A.~B.}\ \bibnamefont
  {Vorontsov}}\ and\ \bibinfo {author} {\bibfnamefont {J.~A.}\ \bibnamefont
  {Sauls}},\ }\bibfield  {title} {\bibinfo {title} {{Thermodynamic properties
  of thin films of superfluid ${}^{3}\mathrm{He}\ensuremath{-}A$}},\ }\href
  {https://doi.org/10.1103/PhysRevB.68.064508} {\bibfield  {journal} {\bibinfo
  {journal} {Phys. Rev. B}\ }\textbf {\bibinfo {volume} {68}},\ \bibinfo
  {pages} {064508} (\bibinfo {year} {2003})}\BibitemShut {NoStop}%
\bibitem [{\citenamefont {Ali}\ \emph {et~al.}(2011)\citenamefont {Ali},
  \citenamefont {Zhang},\ and\ \citenamefont {Sauls}}]{Ali2011Feb}%
  \BibitemOpen
  \bibfield  {author} {\bibinfo {author} {\bibfnamefont {S.}~\bibnamefont
  {Ali}}, \bibinfo {author} {\bibfnamefont {L.}~\bibnamefont {Zhang}},\ and\
  \bibinfo {author} {\bibfnamefont {J.~A.}\ \bibnamefont {Sauls}},\ }\bibfield
  {title} {\bibinfo {title} {{Thermodynamic Potential for Superfluid 3He in
  Aerogel}},\ }\href {https://doi.org/10.1007/s10909-010-0310-4} {\bibfield
  {journal} {\bibinfo  {journal} {J. Low Temp. Phys.}\ }\textbf {\bibinfo
  {volume} {162}},\ \bibinfo {pages} {233} (\bibinfo {year}
  {2011})}\BibitemShut {NoStop}%
\bibitem [{\citenamefont {Zhao}\ \emph {et~al.}(2004)\citenamefont {Zhao},
  \citenamefont {L{\ifmmode\ddot{o}\else\"{o}\fi}fwander},\ and\ \citenamefont
  {Sauls}}]{Zhao2004Oct}%
  \BibitemOpen
  \bibfield  {author} {\bibinfo {author} {\bibfnamefont {E.}~\bibnamefont
  {Zhao}}, \bibinfo {author} {\bibfnamefont {T.}~\bibnamefont
  {L{\ifmmode\ddot{o}\else\"{o}\fi}fwander}},\ and\ \bibinfo {author}
  {\bibfnamefont {J.~A.}\ \bibnamefont {Sauls}},\ }\bibfield  {title} {\bibinfo
  {title} {{Nonequilibrium superconductivity near spin-active interfaces}},\
  }\href {https://doi.org/10.1103/PhysRevB.70.134510} {\bibfield  {journal}
  {\bibinfo  {journal} {Phys. Rev. B}\ }\textbf {\bibinfo {volume} {70}},\
  \bibinfo {pages} {134510} (\bibinfo {year} {2004})}\BibitemShut {NoStop}%
\bibitem [{\citenamefont {Vorontsov}(2009)}]{Vorontsov2009Apr}%
  \BibitemOpen
  \bibfield  {author} {\bibinfo {author} {\bibfnamefont {A.~B.}\ \bibnamefont
  {Vorontsov}},\ }\bibfield  {title} {\bibinfo {title} {{Broken Translational
  and Time-Reversal Symmetry in Unconventional Superconducting Films}},\ }\href
  {https://doi.org/10.1103/PhysRevLett.102.177001} {\bibfield  {journal}
  {\bibinfo  {journal} {Phys. Rev. Lett.}\ }\textbf {\bibinfo {volume} {102}},\
  \bibinfo {pages} {177001} (\bibinfo {year} {2009})}\BibitemShut {NoStop}%
\bibitem [{\citenamefont {Holmvall}\ \emph {et~al.}(2018)\citenamefont
  {Holmvall}, \citenamefont {Vorontsov}, \citenamefont
  {Fogelstr{\ifmmode\ddot{o}\else\"{o}\fi}m},\ and\ \citenamefont
  {L{\ifmmode\ddot{o}\else\"{o}\fi}fwander}}]{Holmvall2018Jun}%
  \BibitemOpen
  \bibfield  {author} {\bibinfo {author} {\bibfnamefont {P.}~\bibnamefont
  {Holmvall}}, \bibinfo {author} {\bibfnamefont {A.~B.}\ \bibnamefont
  {Vorontsov}}, \bibinfo {author} {\bibfnamefont {M.}~\bibnamefont
  {Fogelstr{\ifmmode\ddot{o}\else\"{o}\fi}m}},\ and\ \bibinfo {author}
  {\bibfnamefont {T.}~\bibnamefont {L{\ifmmode\ddot{o}\else\"{o}\fi}fwander}},\
  }\bibfield  {title} {\bibinfo {title} {{Broken translational symmetry at
  edges of high-temperature superconductors}},\ }\href
  {https://doi.org/10.1038/s41467-018-04531-y} {\bibfield  {journal} {\bibinfo
  {journal} {Nat. Commun.}\ }\textbf {\bibinfo {volume} {9}},\ \bibinfo {pages}
  {1} (\bibinfo {year} {2018})}\BibitemShut {NoStop}%
\bibitem [{\citenamefont {Holmvall}\ \emph {et~al.}(2023)\citenamefont
  {Holmvall}, \citenamefont {Wall~Wennerdal}, \citenamefont {H{\aa}kansson},
  \citenamefont {Stadler}, \citenamefont {Shevtsov}, \citenamefont
  {L{\ifmmode\ddot{o}\else\"{o}\fi}fwander},\ and\ \citenamefont
  {Fogelstr{\ifmmode\ddot{o}\else\"{o}\fi}m}}]{Holmvall2023Mar}%
  \BibitemOpen
  \bibfield  {author} {\bibinfo {author} {\bibfnamefont {P.}~\bibnamefont
  {Holmvall}}, \bibinfo {author} {\bibfnamefont {N.}~\bibnamefont
  {Wall~Wennerdal}}, \bibinfo {author} {\bibfnamefont {M.}~\bibnamefont
  {H{\aa}kansson}}, \bibinfo {author} {\bibfnamefont {P.}~\bibnamefont
  {Stadler}}, \bibinfo {author} {\bibfnamefont {O.}~\bibnamefont {Shevtsov}},
  \bibinfo {author} {\bibfnamefont {T.}~\bibnamefont
  {L{\ifmmode\ddot{o}\else\"{o}\fi}fwander}},\ and\ \bibinfo {author}
  {\bibfnamefont {M.}~\bibnamefont
  {Fogelstr{\ifmmode\ddot{o}\else\"{o}\fi}m}},\ }\bibfield  {title} {\bibinfo
  {title} {{SuperConga: An open-source framework for mesoscopic
  superconductivity}},\ }\bibfield  {journal} {\bibinfo  {journal} {Appl. Phys.
  Rev.}\ }\textbf {\bibinfo {volume} {10}},\ \href
  {https://doi.org/10.1063/5.0100324} {10.1063/5.0100324} (\bibinfo {year}
  {2023})\BibitemShut {NoStop}%
\bibitem [{\citenamefont {Tedrow}\ \emph {et~al.}(1984)\citenamefont {Tedrow},
  \citenamefont {Kucera}, \citenamefont {Rainer},\ and\ \citenamefont
  {Orlando}}]{Tedrow1984Apr}%
  \BibitemOpen
  \bibfield  {author} {\bibinfo {author} {\bibfnamefont {P.~M.}\ \bibnamefont
  {Tedrow}}, \bibinfo {author} {\bibfnamefont {J.~T.}\ \bibnamefont {Kucera}},
  \bibinfo {author} {\bibfnamefont {D.}~\bibnamefont {Rainer}},\ and\ \bibinfo
  {author} {\bibfnamefont {T.~P.}\ \bibnamefont {Orlando}},\ }\bibfield
  {title} {\bibinfo {title} {{Spin-Polarized Tunneling Measurement of the
  Antisymmetric Fermi-Liquid Parameter ${G}^{0}$ and Renormalization of the
  Pauli Limiting Field in A1}},\ }\href
  {https://doi.org/10.1103/PhysRevLett.52.1637} {\bibfield  {journal} {\bibinfo
   {journal} {Phys. Rev. Lett.}\ }\textbf {\bibinfo {volume} {52}},\ \bibinfo
  {pages} {1637} (\bibinfo {year} {1984})}\BibitemShut {NoStop}%
\bibitem [{\citenamefont {Alexander}\ \emph {et~al.}(1985)\citenamefont
  {Alexander}, \citenamefont {Orlando}, \citenamefont {Rainer},\ and\
  \citenamefont {Tedrow}}]{Alexander1985May}%
  \BibitemOpen
  \bibfield  {author} {\bibinfo {author} {\bibfnamefont {J.~A.~X.}\
  \bibnamefont {Alexander}}, \bibinfo {author} {\bibfnamefont {T.~P.}\
  \bibnamefont {Orlando}}, \bibinfo {author} {\bibfnamefont {D.}~\bibnamefont
  {Rainer}},\ and\ \bibinfo {author} {\bibfnamefont {P.~M.}\ \bibnamefont
  {Tedrow}},\ }\bibfield  {title} {\bibinfo {title} {{Theory of Fermi-liquid
  effects in high-field tunneling}},\ }\href
  {https://doi.org/10.1103/PhysRevB.31.5811} {\bibfield  {journal} {\bibinfo
  {journal} {Phys. Rev. B}\ }\textbf {\bibinfo {volume} {31}},\ \bibinfo
  {pages} {5811} (\bibinfo {year} {1985})}\BibitemShut {NoStop}%
\bibitem [{\citenamefont {Meservey}\ and\ \citenamefont
  {Tedrow}(1994)}]{Meservey1994Mar}%
  \BibitemOpen
  \bibfield  {author} {\bibinfo {author} {\bibfnamefont {R.}~\bibnamefont
  {Meservey}}\ and\ \bibinfo {author} {\bibfnamefont {P.~M.}\ \bibnamefont
  {Tedrow}},\ }\bibfield  {title} {\bibinfo {title} {{Spin-polarized electron
  tunneling}},\ }\href {https://doi.org/10.1016/0370-1573(94)90105-8}
  {\bibfield  {journal} {\bibinfo  {journal} {Phys. Rep.}\ }\textbf {\bibinfo
  {volume} {238}},\ \bibinfo {pages} {173} (\bibinfo {year}
  {1994})}\BibitemShut {NoStop}%
\bibitem [{\citenamefont {Kashiwaya}\ \emph {et~al.}(2004)\citenamefont
  {Kashiwaya}, \citenamefont {Kashiwaya}, \citenamefont {Prijamboedi},
  \citenamefont {Sawa}, \citenamefont {Kurosawa}, \citenamefont {Tanaka},\ and\
  \citenamefont {Iguchi}}]{Kashiwaya2004Sep}%
  \BibitemOpen
  \bibfield  {author} {\bibinfo {author} {\bibfnamefont {H.}~\bibnamefont
  {Kashiwaya}}, \bibinfo {author} {\bibfnamefont {S.}~\bibnamefont
  {Kashiwaya}}, \bibinfo {author} {\bibfnamefont {B.}~\bibnamefont
  {Prijamboedi}}, \bibinfo {author} {\bibfnamefont {A.}~\bibnamefont {Sawa}},
  \bibinfo {author} {\bibfnamefont {I.}~\bibnamefont {Kurosawa}}, \bibinfo
  {author} {\bibfnamefont {Y.}~\bibnamefont {Tanaka}},\ and\ \bibinfo {author}
  {\bibfnamefont {I.}~\bibnamefont {Iguchi}},\ }\bibfield  {title} {\bibinfo
  {title} {{Anomalous magnetic-field tunneling of
  $\mathrm{Y}{\mathrm{Ba}}_{2}{\mathrm{Cu}}_{3}{\mathrm{O}}_{7\ensuremath{-}\ensuremath{\delta}}$
  junctions: Possible detection of non-Fermi-liquid states}},\ }\href
  {https://doi.org/10.1103/PhysRevB.70.094501} {\bibfield  {journal} {\bibinfo
  {journal} {Phys. Rev. B}\ }\textbf {\bibinfo {volume} {70}},\ \bibinfo
  {pages} {094501} (\bibinfo {year} {2004})}\BibitemShut {NoStop}%
\bibitem [{\citenamefont {Matsubara}\ and\ \citenamefont
  {Kontani}(2020{\natexlab{a}})}]{Matsubara2020Feb}%
  \BibitemOpen
  \bibfield  {author} {\bibinfo {author} {\bibfnamefont {S.}~\bibnamefont
  {Matsubara}}\ and\ \bibinfo {author} {\bibfnamefont {H.}~\bibnamefont
  {Kontani}},\ }\bibfield  {title} {\bibinfo {title} {{Emergence of strongly
  correlated electronic states driven by the Andreev bound state in $d$-wave
  superconductors}},\ }\href {https://doi.org/10.1103/PhysRevB.101.075114}
  {\bibfield  {journal} {\bibinfo  {journal} {Phys. Rev. B}\ }\textbf {\bibinfo
  {volume} {101}},\ \bibinfo {pages} {075114} (\bibinfo {year}
  {2020}{\natexlab{a}})}\BibitemShut {NoStop}%
\bibitem [{\citenamefont {Matsubara}\ and\ \citenamefont
  {Kontani}(2020{\natexlab{b}})}]{Matsubara2020Jun}%
  \BibitemOpen
  \bibfield  {author} {\bibinfo {author} {\bibfnamefont {S.}~\bibnamefont
  {Matsubara}}\ and\ \bibinfo {author} {\bibfnamefont {H.}~\bibnamefont
  {Kontani}},\ }\bibfield  {title} {\bibinfo {title} {{Emergence of
  $d\ifmmode\pm\else\textpm\fi{}ip$-wave superconducting state at the edge of
  $d$-wave superconductors mediated by ferromagnetic fluctuations driven by
  Andreev bound states}},\ }\href {https://doi.org/10.1103/PhysRevB.101.235103}
  {\bibfield  {journal} {\bibinfo  {journal} {Phys. Rev. B}\ }\textbf {\bibinfo
  {volume} {101}},\ \bibinfo {pages} {235103} (\bibinfo {year}
  {2020}{\natexlab{b}})}\BibitemShut {NoStop}%
\bibitem [{\citenamefont {Arndt}\ \emph {et~al.}(2023)\citenamefont {Arndt},
  \citenamefont {Bangerth}, \citenamefont {Bergbauer}, \citenamefont {Feder},
  \citenamefont {Fehling}, \citenamefont {Heinz}, \citenamefont {Heister},
  \citenamefont {Heltai}, \citenamefont {Kronbichler}, \citenamefont {Maier},
  \citenamefont {Munch}, \citenamefont {Pelteret}, \citenamefont {Turcksin},
  \citenamefont {Wells},\ and\ \citenamefont {Zampini}}]{dealII95}%
  \BibitemOpen
  \bibfield  {author} {\bibinfo {author} {\bibfnamefont {D.}~\bibnamefont
  {Arndt}}, \bibinfo {author} {\bibfnamefont {W.}~\bibnamefont {Bangerth}},
  \bibinfo {author} {\bibfnamefont {M.}~\bibnamefont {Bergbauer}}, \bibinfo
  {author} {\bibfnamefont {M.}~\bibnamefont {Feder}}, \bibinfo {author}
  {\bibfnamefont {M.}~\bibnamefont {Fehling}}, \bibinfo {author} {\bibfnamefont
  {J.}~\bibnamefont {Heinz}}, \bibinfo {author} {\bibfnamefont
  {T.}~\bibnamefont {Heister}}, \bibinfo {author} {\bibfnamefont
  {L.}~\bibnamefont {Heltai}}, \bibinfo {author} {\bibfnamefont
  {M.}~\bibnamefont {Kronbichler}}, \bibinfo {author} {\bibfnamefont
  {M.}~\bibnamefont {Maier}}, \bibinfo {author} {\bibfnamefont
  {P.}~\bibnamefont {Munch}}, \bibinfo {author} {\bibfnamefont {J.-P.}\
  \bibnamefont {Pelteret}}, \bibinfo {author} {\bibfnamefont {B.}~\bibnamefont
  {Turcksin}}, \bibinfo {author} {\bibfnamefont {D.}~\bibnamefont {Wells}},\
  and\ \bibinfo {author} {\bibfnamefont {S.}~\bibnamefont {Zampini}},\
  }\bibfield  {title} {\bibinfo {title} {The \texttt{deal.II} library, version
  9.5},\ }\href {https://doi.org/10.1515/jnma-2023-0089} {\bibfield  {journal}
  {\bibinfo  {journal} {Journal of Numerical Mathematics}\ }\textbf {\bibinfo
  {volume} {31}},\ \bibinfo {pages} {231} (\bibinfo {year} {2023})}\BibitemShut
  {NoStop}%
\bibitem [{\citenamefont {Cockburn}\ \emph {et~al.}(2000)\citenamefont
  {Cockburn}, \citenamefont {Karniadakis},\ and\ \citenamefont
  {Shu}}]{Cockburn2000}%
  \BibitemOpen
  \bibfield  {author} {\bibinfo {author} {\bibfnamefont {B.}~\bibnamefont
  {Cockburn}}, \bibinfo {author} {\bibfnamefont {G.~E.}\ \bibnamefont
  {Karniadakis}},\ and\ \bibinfo {author} {\bibfnamefont {C.-W.}\ \bibnamefont
  {Shu}},\ }\bibfield  {title} {\bibinfo {title} {{The Development of
  Discontinuous Galerkin Methods}},\ }in\ \href
  {https://doi.org/10.1007/978-3-642-59721-3_1} {\emph {\bibinfo {booktitle}
  {{Discontinuous Galerkin Methods}}}}\ (\bibinfo  {publisher} {Springer},\
  \bibinfo {address} {Berlin, Germany},\ \bibinfo {year} {2000})\ pp.\ \bibinfo
  {pages} {3--50}\BibitemShut {NoStop}%
\bibitem [{\citenamefont {Cockburn}(2003)}]{Cockburn2003Nov}%
  \BibitemOpen
  \bibfield  {author} {\bibinfo {author} {\bibfnamefont {B.}~\bibnamefont
  {Cockburn}},\ }\bibfield  {title} {\bibinfo {title} {{Discontinuous Galerkin
  methods}},\ }\href {https://doi.org/10.1002/zamm.200310088} {\bibfield
  {journal} {\bibinfo  {journal} {Z. angew. Math. Mech.}\ }\textbf {\bibinfo
  {volume} {83}},\ \bibinfo {pages} {731} (\bibinfo {year} {2003})}\BibitemShut
  {NoStop}%
\bibitem [{\citenamefont {Arnold}\ \emph {et~al.}(2002)\citenamefont {Arnold},
  \citenamefont {Brezzi}, \citenamefont {Cockburn},\ and\ \citenamefont
  {Marini}}]{Arnold2002}%
  \BibitemOpen
  \bibfield  {author} {\bibinfo {author} {\bibfnamefont {D.~N.}\ \bibnamefont
  {Arnold}}, \bibinfo {author} {\bibfnamefont {F.}~\bibnamefont {Brezzi}},
  \bibinfo {author} {\bibfnamefont {B.}~\bibnamefont {Cockburn}},\ and\
  \bibinfo {author} {\bibfnamefont {L.~D.}\ \bibnamefont {Marini}},\ }\bibfield
   {title} {\bibinfo {title} {{Unified Analysis of Discontinuous Galerkin
  Methods for Elliptic Problems}},\ }\href
  {https://epubs.siam.org/doi/10.1137/S0036142901384162} {\bibfield  {journal}
  {\bibinfo  {journal} {SIAM J. Numer. Anal.}\ }\textbf {\bibinfo {volume}
  {39}},\ \bibinfo {pages} {1749–1779} (\bibinfo {year} {2002})}\BibitemShut
  {NoStop}%
\bibitem [{\citenamefont {Brezzi}\ \emph {et~al.}(2004)\citenamefont {Brezzi},
  \citenamefont {Marini},\ and\ \citenamefont
  {S{\ifmmode\ddot{u}\else\"{u}\fi}li}}]{Brezzi2004Dec}%
  \BibitemOpen
  \bibfield  {author} {\bibinfo {author} {\bibfnamefont {F.}~\bibnamefont
  {Brezzi}}, \bibinfo {author} {\bibfnamefont {L.~D.}\ \bibnamefont {Marini}},\
  and\ \bibinfo {author} {\bibfnamefont {E.}~\bibnamefont
  {S{\ifmmode\ddot{u}\else\"{u}\fi}li}},\ }\bibfield  {title} {\bibinfo {title}
  {{Discontinuous Galerkin Methods for First-Order Hyperbolic Problems}},\
  }\href {https://doi.org/10.1142/S0218202504003866} {\bibfield  {journal}
  {\bibinfo  {journal} {Math. Models Methods Appl. Sci.}\ }\textbf {\bibinfo
  {volume} {14}},\ \bibinfo {pages} {1893} (\bibinfo {year}
  {2004})}\BibitemShut {NoStop}%
\bibitem [{\citenamefont {Grein}\ \emph {et~al.}(2013)\citenamefont {Grein},
  \citenamefont {L{\ifmmode\ddot{o}\else\"{o}\fi}fwander},\ and\ \citenamefont
  {Eschrig}}]{Grein2013Aug}%
  \BibitemOpen
  \bibfield  {author} {\bibinfo {author} {\bibfnamefont {R.}~\bibnamefont
  {Grein}}, \bibinfo {author} {\bibfnamefont {T.}~\bibnamefont
  {L{\ifmmode\ddot{o}\else\"{o}\fi}fwander}},\ and\ \bibinfo {author}
  {\bibfnamefont {M.}~\bibnamefont {Eschrig}},\ }\bibfield  {title} {\bibinfo
  {title} {{Inverse proximity effect and influence of disorder on triplet
  supercurrents in strongly spin-polarized ferromagnets}},\ }\href
  {https://doi.org/10.1103/PhysRevB.88.054502} {\bibfield  {journal} {\bibinfo
  {journal} {Phys. Rev. B}\ }\textbf {\bibinfo {volume} {88}},\ \bibinfo
  {pages} {054502} (\bibinfo {year} {2013})}\BibitemShut {NoStop}%
\bibitem [{\citenamefont {Johnson}(2009)}]{Johnson2009Jan}%
  \BibitemOpen
  \bibfield  {author} {\bibinfo {author} {\bibfnamefont {C.}~\bibnamefont
  {Johnson}},\ }\href@noop {} {\emph {\bibinfo {title} {{Numerical Solution of
  Partial Differential Equations by the Finite Element Method (Dover Books on
  Mathematics)}}}}\ (\bibinfo  {publisher} {Dover Publications},\ \bibinfo
  {year} {2009})\BibitemShut {NoStop}%
\bibitem [{\citenamefont {Ozaki}(2007)}]{Ozaki2007Jan}%
  \BibitemOpen
  \bibfield  {author} {\bibinfo {author} {\bibfnamefont {T.}~\bibnamefont
  {Ozaki}},\ }\bibfield  {title} {\bibinfo {title} {{Continued fraction
  representation of the Fermi-Dirac function for large-scale electronic
  structure calculations}},\ }\href
  {https://doi.org/10.1103/PhysRevB.75.035123} {\bibfield  {journal} {\bibinfo
  {journal} {Phys. Rev. B}\ }\textbf {\bibinfo {volume} {75}},\ \bibinfo
  {pages} {035123} (\bibinfo {year} {2007})}\BibitemShut {NoStop}%
\end{thebibliography}%

\end{document}